\documentclass[useAMS,usenatbib]{mn2e}
\usepackage{times}
\usepackage{amssymb}
\usepackage{psfig}
\usepackage{graphicx}
\usepackage{rotating}
\usepackage{lscape}
\usepackage{multirow}
\usepackage{fixltx2e}
\usepackage{color}

\begin{document}

% If your system does not have the AMS fonts version 2.0 installed, then
% remove the useAMS option.
%
% useAMS allows you to obtain upright Greek characters.
% e.g. \umu, \upi etc.  See the section on "Upright Greek characters" in
% this guide for further information.
%
% If you are using AMS 2.0 fonts, bold math letters/symbols are available
% at a larger range of sizes for NFSS release 1 and 2 (using \boldmath or
% preferably \bmath).
%
% The usenatbib command allows the use of Patrick Daly's natbib.sty for
% cross-referencing.
%
% If you wish to typeset the paper in Times font (if you do not have the
% PostScript Type 1 Computer Modern fonts you will need to do this to get
% smoother fonts in a PDF file) then uncomment the next line
% \usepackage{Times}

%%%%% AUTHORS - PLACE YOUR OWN MACROS HERE %%%%%

%%%%%%%%%%%%%%%%%%%%%%%%%%%%%%%%%%%%%%%%%%%%%%%%

\title[KMOS-HiZELS metallicity gradients]{A relationship between specific star formation rate and metallicity gradient within \boldmath$z\sim1$ galaxies from KMOS-HiZELS} 

\author[J.P. Stott et al.]{John P. Stott$^{1}$\thanks{E-mail: j.p.stott@durham.ac.uk}, David Sobral$^{2,3}$, A. M. Swinbank$^{1}$, Ian Smail$^{1}$, Richard Bower$^{1}$, \newauthor Philip N. Best$^{4}$,    Ray M. Sharples$^{1}$, James E. Geach$^{5}$, Jorryt Matthee$^{3}$ \\
$^{1}$ Institute for Computational Cosmology, Durham University, South Road, Durham, DH1 3LE, UK\\
$^{2}$ Centro de Astronomia e Astrof\'{\i}sica da Universidade de Lisboa, Observat\'{o}rio Astron\'{o}mico de Lisboa, Tapada da Ajuda, 1349-018 Lisboa, Portugal\\
$^{3}$ Leiden Observatory, Leiden University, P.O. Box 9513, NL-2300 RA Leiden, The Netherlands\\
$^{4}$ SUPA, Institute for Astronomy, Royal Observatory of Edinburgh, Blackford Hill, Edinburgh, EH9 3HJ, UK\\
$^{5}$ Centre for Astrophysics Research, Science \& Technology Research Institute, University of Hertfordshire, Hatfield, AL10 9AB \\
}

\date{}

\pagerange{\pageref{firstpage}--\pageref{lastpage}} \pubyear{2013}

\maketitle

\label{firstpage}

\begin{abstract}

We have observed a sample of typical $z\sim1$ star forming galaxies, selected from the HiZELS survey, with the new KMOS near-infrared, multi-IFU instrument on the VLT, in order to obtain their dynamics and metallicity gradients. The majority of our galaxies have a metallicity gradient consistent with being flat or negative (i.e. higher metallicity cores than outskirts). Intriguingly, we find a trend between metallicity gradient and specific star formation rate (sSFR), such that galaxies with a high sSFR tend to have relatively metal-poor centres, a result which is strengthened when combined with datasets from the literature. This result appears to explain the discrepancies reported between different high redshift studies and varying claims for evolution. From a galaxy evolution perspective, the trend we see would mean that a galaxy's sSFR is governed by the amount of metal poor gas that can be funnelled into its core, triggered either by merging or through efficient accretion. In fact merging may play a significant role as it is the starburst galaxies at all epochs, which have the more positive metallicity gradients. Our results may help to explain the origin of the fundamental metallicity relation, in which galaxies at a fixed mass are observed to have lower metallicities at higher star formation rates, especially if the metallicity is measured in an aperture encompassing only the central regions of the galaxy. Finally, we note that this study demonstrates the power of KMOS as an efficient instrument for large scale resolved galaxy surveys. 

\end{abstract}


\section{Introduction}
\label{sec:intro}
The gas phase metallicity of a galaxy reflects the past star-forming activity and the history of both gas inflow and outflow of the system. Observations of galaxy metallicity and its dependence on mass can therefore be used to trace this history by comparing local samples  (e.g. \citealt{tremonti2004,kewley2008}) to those at higher redshifts (e.g. \citealt{savaglio2005,erb2006, maiolino2008,lamareille2009,perezm2009,yabe2012,zahid2013,stott2013b}). The results of such studies have generally found a strong evolution in the gas phase metallicity, with galaxies being more metal-poor at increasing redshift. However, this is perhaps because the observed high redshift galaxies tend to be more highly star forming, as there is now evidence that galaxies are found to sit on a similar plane of mass, metallicity and star formation rate (SFR) in both the local and high redshift Universe \citep{mannucci2010,laralopez2010,stott2013b}. 

As well as studying the individual and average metallicities within galaxy populations, more detailed observations that trace the variation of metallicity within galaxies can also be employed in order to understand their evolution. Due to the spatially resolved, high signal-to-noise spectral observations required, this has mainly been performed for relatively small galaxy samples in the local Universe. In the first comprehensive study, \cite{searle1971} measured the differences in line ratios, and therefore the implied chemical abundance, between different HII regions within the same galaxy. The key result of \cite{searle1971} and subsequent studies is that galaxies in the local Universe tend to have negative radial metallicity gradients, such that the stars and gas in the outer regions appear less metal rich than those in the centre (e.g. \citealt{shields1974,mccall1985,vilacostas1992,zaritsky1994,garnett1997,vanzee1998,bresolin2012,sanchez2012,sanchez2014} and see the review by \citealt{henry1999}). At high redshift detailed observations become more challenging and the results more contradictory with some authors finding abundance gradients that are consistent with being flat or negative \citep{swinbank2012,jones2013} and others seeing evidence for positive gradients \citep{cresci2010, queyrel2012}. 

From a theoretical perspective, disc galaxies that follow inside-out growth tend to have initially steep negative abundance gradients which then flatten at later times (e.g. \citealt{marcon2010,stinson2010,gibson2013}). Observational support for this has been claimed, as \cite{jones2013} find a small subset of their $z=2$ galaxies possess significantly steeper negative abundance gradients than local galaxies. Simulations also show that merging events will rapidly flatten existing metallicity gradients of galaxies by inducing an inflow of metal-poor gas to their central regions  \citep{rupke2010a}. This effect has been witnessed in observations of low redshift interacting galaxies \citep{rupke2010b}. There is also the possibility that so-called `cold flows' of metal-poor gas at high redshift could lead to lower central metallicities (e.g. \citealt{cresci2010}) but there is some uncertainty as to how and where any inflowing material is deposited \citep{keres2005,dekel2009}.

In order to study abundance gradients at $z\sim1$, we have observed a representative sample of star forming galaxies with the K-band Multi-Object Spectrograph (KMOS), which is a near-infrared multiple integral field spectrograph \citep{sharples2013}. These galaxies are drawn from our large (10-square degree) narrow-band H$\alpha$ survey in SA22 using WIRcam/CFHT (CF-HiZELS, Sobral et al. in prep, \citealt{sobral2013kmos}, \citealt{matthee2014}). Due to the depth achieved by our observations ($\sim$0.2\,L$^*_{z=0.8}$), the majority of our targets are `typical' galaxies at this epoch which will likely evolve into $\sim$\,L$^{\star}$ (or SFR$^*$) galaxies by \emph{z}\,=\,0. This survey builds on our previous H$\alpha$ narrow-band imaging of degree-sized areas in redshift slices at $z$\,=\,0.40, 0.84, 1.47 and 2.23 from HiZELS  \citep[][]{geach2008,sobral2009,sobral2012,sobral2013}.  

The KMOS observations were performed as part of the Science Verification (SV\footnote{http://www.eso.org/sci/activities/vltsv/kmossv.html}) and focus on two relatively over-dense regions of H$\alpha$ emitters within SA22 field of the CF-HiZELS survey. These observations provide spatially resolved H$\alpha$ and [NII] measurements which allow us to obtain resolved dynamics and metallicities. The dynamical results for the first region are presented in \cite{sobral2013kmos}. In this paper, we use the combined dataset from both regions to investigate the chemical abundance gradients and the dynamical properties of the galaxies.  

We use a cosmology with $\Omega_{\Lambda}$\,=\,0.73, $\Omega_{m}$\,=\,0.27, and H$_{0}$\,=\,72\,km\,s$^{-1}$\,Mpc$^{-1}$. %It's turtles all the way down.
We note that $1''$ corresponds to 7.6\,kpc at $z=0.81$ (the median redshift of the galaxies presented in this paper). All quoted magnitudes are on the AB system and we use a \cite{chabrier2003} IMF throughout.

\section{Sample and Data}
Our targets are selected from a survey using the narrow-band (NB) lowOH2 filter ($\lambda=1187\pm5$\,nm) on WIRCam\,/\,CFHT \citep{puget2004}, which covers a 10\,deg$^2$ contiguous area in SA22 (\citealt{sobral2013kmos}; Sobral et al. in prep). The survey yields $\sim3000$ robust H$\alpha$ emitters at $z$\,=\,0.81\,$\pm$\,0.01 (see \citealt{sobral2013} and Sobral et al. in prep for details on the spectroscopic and photometric redshifts, and colour-colour selection). We use the wealth of ancillary data, including 7-band photometric coverage (from $u$ to $K$-band, available in this field from the CFHT Legacy Survey [CFHTLS] and the UKIRT Infrared Deep Sky Survey [UKIDSS]) to compute stellar masses for all of the H$\alpha$ emitters in the parent sample following \cite{sobral2011,sobral2014}. The two regions that we target with KMOS are at R.A. 22\,19\,30.3,  Dec. $+$00\,38\,59 and R.A. 22\,19\,41.5, Dec. $+$00\,23\,20 (J2000). Both of these observations were taken as part of the KMOS SV and so we label the two fields KMOS-HiZELS-SV1 (see also \citealt{sobral2013kmos}) and KMOS-HiZELS-SV2 (this paper) respectively.

\begin{figure*}
   \centering

\includegraphics[scale=0.13, trim=0 0 0 0, clip=true]{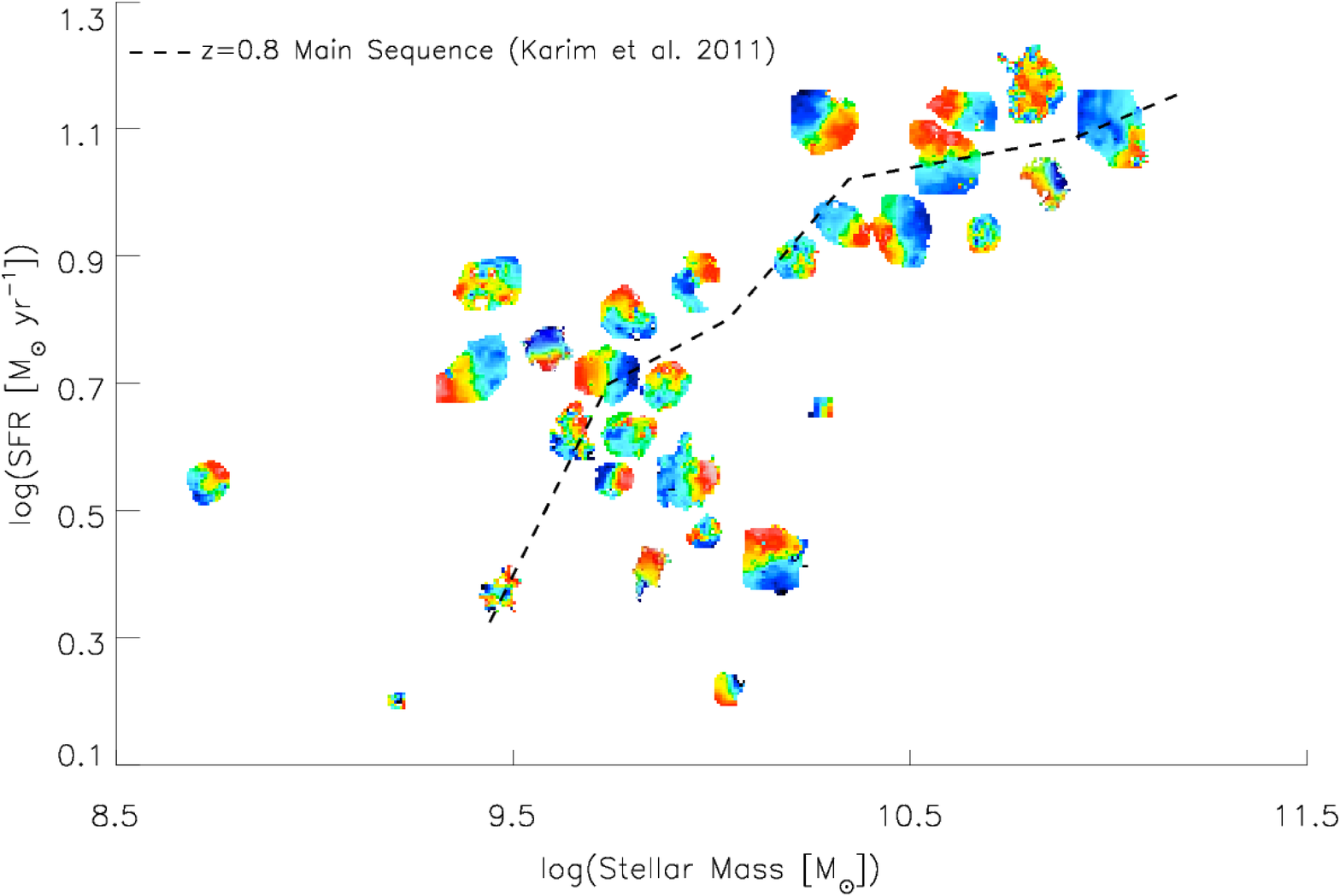}

\caption[]{The SFR plotted against stellar mass for the 29 resolved galaxies in the KMOS-HiZELS sample with the data points represented by their velocity fields (normalised to their maximum observed velocities to make the rotation visible for a range of rotation speeds). For the velocity fields, red denotes a positive (recessional) velocity relative to the systemic redshift (green), while blue is negative. Note, positions are approximate to avoid galaxy velocity fields from overlapping. The dashed line represents the location of the `main sequence' of star forming galaxies at $z=0.8-1.0$ from \cite{karim2011}, demonstrating that our sample is typical for this epoch. }
   \label{fig:ms}
\end{figure*}

The KMOS spectrograph consists of 24 integral field units (IFUs) that patrol a 7.2\,arcminute field. Each IFU has an area of 2.8$''\times2.8''$ with $0.2''\times0.2''$ spatial pixels. We identified target H$\alpha$ emitters with narrow-band H$\alpha$ fluxes brighter than 1\,$\times$\,10$^{-16}$\,erg\,s$^{-1}$\,cm$^{-2}$, (star formation rates $>3.5$\,M$_{\odot}$\,yr$^{-1}$, assuming 1 mag of extinction, \citealt{kennicutt1998} and a \citealt{chabrier2003} IMF) which lie within 7$'$ diameter regions centred on KMOS-HiZELS-SV1 and KMOS-HiZELS-SV2. For KMOS-HiZELS-SV1 we selected the 21 of these galaxies which were brighter than $K_{\rm AB}$\,$\sim$\,21.5 (roughly corresponding to stellar mass $M_{\star}> $\,10$^{9.75}$\,$M_{\odot}$). KMOS-HiZELS-SV2 has a lower number density of HiZELS sources and so we selected 12 $\rm H\alpha$ emitters and a further 9 galaxies with spectroscopic redshifts in the range $0.8<z<1.0$ from the VIMOS/VLT Deep Survey (VVDS) survey \citep{lefevre2005}. We therefore selected 42 galaxies for observations during science verification time with KMOS (although only 39 were observed due to technical problems, see below). The galaxies in this KMOS sample have a median stellar mass of $\sim10^{10.1}$\,M$_{\odot}$\,yr$^{-1}$, a median SFR of 7\,M$_{\odot}$\,yr$^{-1}$ and a median sSFR (SFR/M$_\star$) of 0.5\,Gyr$^{-1}$ (see Fig. ~\ref{fig:ms}). Our KMOS sources are typical star-forming galaxies at their redshift (2--14\,M$_{\odot}$\,yr$^{-1}$, while the characteristic SFR [SFR$^*$] at $z\sim0.8$ is $\sim8$\,M$_{\odot}$\,yr$^{-1}$, c.f. \citealt{sobral2014}).

KMOS observations were taken in 2013 on June 29, July 1 (KMOS-HiZELS-SV1) and September 25 (KMOS-HiZELS-SV2).  During the observations the average $J$-band seeing was approximately 0.7$''$. We used the $YJ$-band grating in order to cover the H$\alpha$ emission, which at $z\sim$\,0.81 (CF-HiZELS narrow band) is redshifted to $\sim$\,1.187$\mu$m. In this configuration, the spectral resolution is R\,=\,$\lambda$\,/\,$\Delta\lambda\,\sim\,3400$. We also deployed three IFUs (one per KMOS spectrograph) to (blank) sky positions to improve the sky-subtraction during the data reduction. Observations were carried out using an ABA (object-sky-object) sequence, with 450s integration per position, in which we chopped by 5$''$ to sky, and each observation was dithered by up to 0.2$''$. The total on-source integration time was 1.25\,hrs per galaxy. During the KMOS-HiZELS-SV1 observations, three of the IFUs were disabled and so only 18 galaxies were observed in this pointing making 39 in total.

To reduce the data, we used the {\sc esorex}\,/\,{\sc spark} pipeline \citep{davies2013}, which extracts the slices from each IFU, flatfields and wavelength calibrates the data to form a datacube. We reduced each AB pair separately, and improved the sky OH subtraction in each AB pair for each IFU using the data from the sky IFU from the appropriate spectrograph (using the sky-subtraction techniques described in \citealt{davies2007}). We then combined the data into the final datacube using a clipped average. We note that both the effects of instrumental resolution and the spatial PSF are taken into account throughout the analysis and included in the error estimation.

For the KMOS-HiZELS-SV2 observations, two of the 21 galaxies observed returned no evidence of an emission line and a further six did not have resolved $\rm H\alpha$ emission (three of which came from the VVDS selection rather than the CF-HiZELS narrow-band). For the remainder of this paper we concentrate on the 29 resolved galaxies from the two KMOS-HiZELS pointings.

\section{Analysis \& Results}

\subsection{Dynamics}
\label{sec:dyn}
We begin by determining the dynamical properties of the KMOS-HiZELS galaxies via disc model fitting and kinemetry before studying their resolved metallicities (see also \citealt{sobral2013kmos}).

We collapse each reduced datacube into a one dimensional spectrum and measure the redshift by fitting a Gaussian profile to the H$\alpha$ and [NII] emission lines, for which we also recover their total flux. To measure the H$\alpha$ dynamics of each galaxy, we fit the H$\alpha$ and [NII] emission lines spaxel-to-spaxel using a $\chi^{2}$ minimisation procedure (accounting for the increased noise at the positions of the sky lines). We initially try to identify the H$\alpha$ line in a 0.4$''$\,$\times$\,0.4$''$ region ($\sim$\,3\,kpc), and if the fit fails to detect the line with a signal-to-noise $>$\,5, the region is increased to 0.6$''$\,$\times$\,0.6$''$. When this criterion is met the H$\alpha$ and [NII] emission lines are fitted allowing the centroid, intensity and width of the Gaussian profile to find their optimum values (the FWHM of the H$\alpha$ and [NII] lines are coupled in the fit).  Uncertainties are then calculated by perturbing each parameter, one at a time, allowing the remaining parameters to find their optimum values, until $\Delta\chi^2$\,=\,1 is reached.

The measured velocity fields for the resolved KMOS-HiZELS galaxies are displayed in Fig. \ref{fig:ms} at the approximate positions of their stellar mass and SFR. Fig.~\ref{fig:ms} is therefore a plot of the so called star forming `main sequence' \citep{noeske2007}, with the points represented by the galaxy velocity fields. The position of the star forming main sequence at $z=0.8-1$ from  \cite{karim2011} is included, demonstrating that these galaxies are typical star forming systems at this epoch. The majority of the galaxies display velocity gradients in their dynamics, with observed peak-to-peak differences ranging from $\Delta v$\,$\sim$\,40--300\,km\,s$^{-1}$.  

Many of these galaxies have H$\alpha$ velocity fields which resemble rotating systems (characteristic `spider' patterns in the velocity fields and line of sight velocity dispersion profiles which peak near the central regions).  Therefore, we attempt to model the two dimensional velocity field to identify the dynamical centre and kinematic major axis.  We follow \citet{swinbank2012} to construct two dimensional models with an input rotation curve following an {\rm arctan} function [$v(r)$\,=\,$\frac{2}{\pi}$\,$v_{\rm asym}$\,${\rm arctan}(r/r_t)$], where $v_{\rm asym}$ is the asymptotic rotational velocity and $r_{\rm t}$ is the effective radius at which the rotation curve turns over \citep{courteau1997}.  The suite of two dimensional models which we fit to the data have six free parameters ([x,y] centre, position angle (PA), $r_{\rm t}$, $v_{\rm asym}$, and disc inclination) and we use a genetic algorithm \citep{charbonneau1995} to find the best model \citep[see][]{swinbank2012}. 

The best fit dynamical model produces a dynamical centre and position angle of the disc allowing us to extract the one dimensional rotation curve and velocity dispersion profiles from the major kinematic axis of each galaxy. Despite the relatively short integration time (1.25\,hrs on source), the data yield clear rotation curves which turn over (or flatten) for at least ten of these galaxies (see also \citealt{sobral2013kmos}). 

We also measure the effective radii ($r_{e}$) of the KMOS-HiZELS sample by fitting a 2-dimensional S\'{e}rsic profile to UKIDSS $K$-band images of the galaxies using the {\sc galfit} (version 3) software package \citep{peng2002}. This software requires reasonable initial input parameters such as position, apparent magnitude and ellipticity, all of which are estimated by first running the {\sc sextractor} package \citep{bertin1996} so that the iterative fitting process converges to the correct solution in the shortest possible time. {\sc galfit} deconvolves the atmospheric seeing for this ground-based imaging.

As discussed in \S\ref{sec:intro},  it is likely that merging events and/or cold flows may affect the metallicity gradient of galaxies by transporting gas of differing abundances throughout the disc \citep{cresci2009,rupke2010a}. In order to study the presence of mergers in our sample we perform further analysis, to distinguish between a galaxy with dynamics dominated by ordered rotation or by disturbed kinematics via a method known as `kinemetry'. Kinemetry measures the asymmetry of the velocity field and spatially resolved line-of-sight velocity dispersion for each galaxy. This method has been well calibrated and tested at low redshift \citep[e.g.\ ][]{krajnovic2006}, and used at high redshift to determine the strength of deviations of the observed velocity and dispersion maps from an ideal rotating disc (\citealt{shapiro2008}; \citealt{swinbank2012}; \citealt{sobral2013kmos}; but see also \citealt{goncalves2010}).  Briefly, in this modelling, the velocity and velocity dispersion maps are described by a series of concentric ellipses of increasing semi-major axis length, as defined by the system centre, position angle and inclination. Along each ellipse, the moment map as a function of angle is extracted and decomposed into its Fourier series which have coefficients $k_n$ at each radii (see \citealt{krajnovic2006} for more details).

We measure the velocity field and velocity dispersion asymmetry (K$_{\rm V}$ and K$_{\sigma}$ respectively) for all of the galaxies in our sample. For an ideal disc, the values of K$_{\rm V}$ and K$_{\sigma}$ will be zero. In contrast, in a merging system, strong deviations from the idealised case causes large values of K$_{\rm V}$ and K$_{\sigma}$ (which can reach K$_{\rm V}\sim$\,K$_{\sigma}\sim$\,10 for very disturbed systems).  The total asymmetry, K$_{\rm Tot}$ is K$_{\rm Tot}^2$\,=\,K$_{\rm  V}^2$\,+\,K$_{\sigma}^2$. The majority of our sample have K$_{\rm Tot}\lesssim0.5$ and are therefore consistent with being rotation dominated discs, despite residing in a relatively over-dense region, which may in general lead to an increased merger rate. {The number of galaxies with K$_{\rm Tot}>0.5$ is four, which is consistent with the 10\% merger fraction found on the main sequence at $z\sim0.8$ by \cite{stott2013}}. The dynamical properties of the KMOS-HiZELS sample are given in Table \ref{tab:samp}.

\begin{table*}
\begin{center}
\caption[]{The details of the KMOS-HiZELS sample. The CF-HiZELS galaxies are named CFHT-NBJ and the VVDS galaxies are numbered by our own internal catalogue system. The $v_{80}$ parameter is the inclination corrected rotation speed at $r_{80}$ ($r_{80}=2.2r_e$). The KMOS-HiZELS-SV1 sample data are presented in \cite{sobral2013kmos}, and should be cited as such, but are included here below the horizontal line for completeness.}

\label{tab:samp}
\small\begin{tabular}{lclcccrrllrcc}
%\hline
\hline
Galaxy&R.A.&Dec.&$z$&$K_{\rm AB}$&$r_e$ &$\rm [NII]/H\alpha$& $\rm \log(M_\star)$ &SFR&$v_{80}$&K$_{\rm Tot}$\\%$\rm \log(M_{\star}/M_{\odot})$ &metgrad \\
&\multicolumn{2}{c}{[J2000]}&&&(kpc)&&[$\rm M_\odot$]&[$\rm M_\odot\,yr^{-1}$]&[km\,s$^{-1}$]\\
%&\multicolumn{2}{c}{(J2000)}&&(kpc) &(kpc)&\\
\hline
%KMOS-HiZELS-SV2&&&&&&&&&\\
CFHT-NBJ-C339&22:19:46.96&+00:25:02.5&0.8135&20.12&3.0&$1.28\pm0.12$&$10.6\pm0.1$&11.0&146.&$0.5\pm0.5$\\
CFHT-NBJ-C343&22:19:48.65&+00:21:28.4&0.8100&20.85&4.7&$0.32\pm0.13$&$10.5\pm0.2$&4.1&224.&$0.3\pm0.1$\\
CFHT-NBJ-956&22:19:27.05&+00:23:42.4&0.8095&21.43&4.5&$0.15\pm0.28$&$10.1\pm0.2$&4.1&231.&$0.2\pm0.1$\\
CFHT-NBJ-1209&22:19:40.16&+00:22:38.5&0.8085&21.76&10.4&$0.13\pm0.41$&$9.4\pm0.1$&5.4&219.&$0.1\pm0.7$\\
CFHT-NBJ-1478&22:19:41.06&+00:22:34.2&0.8105&22.10&3.9&$0.18\pm0.27$&$9.9\pm0.4$&4.6&148.&$5.1\pm0.2$\\
CFHT-NBJ-2044&22:19:34.37&+00:23:00.4&0.8099&19.67&8.3&$0.59\pm0.16$&$11.0\pm0.1$&12.5&260.&$0.2\pm0.1$\\
CFHT-NBJ-2048&22:19:51.67&+00:21:00.9&0.8155&22.90&5.8&$0.11\pm0.36$&$8.8\pm0.1$&3.5&89.&$0.3\pm1.1$\\
VVDS-432&22:19:46.70&+00:21:35.4&0.8095&21.24&4.8& $0.17\pm0.53$&$10.1\pm0.2$&1.2&144.& ...\\
VVDS-503&22:19:51.16&+00:25:42.2&0.9925&21.82&4.2&$0.19\pm0.21$&$9.4\pm0.1$&7.6&62.& ...\\
VVDS-588&22:19:32.41&+00:21:01.0&0.8770&20.90&2.2& $0.54\pm0.19$&$10.1\pm0.1$&2.2&207.&$0.5\pm0.7$\\
VVDS-888&22:19:38.00&+00:20:07.4&0.8331&22.10&1.3&$0.27\pm0.15$&$9.7\pm0.1$&4.6&56.&$0.4\pm9.2$\\
VVDS-942&22:19:39.44&+00:25:29.3&0.8095&23.41&4.0&... &$9.2\pm0.4$&1.6&132.& ...\\
VVDS-944&22:19:39.73&+00:24:02.4&0.8970&22.31&2.1& ...&$9.5\pm0.2$&2.3&258.&$0.9\pm0.3$\\
\hline
\multicolumn{3}{l}{KMOS-HiZELS-SV1, from Sobral et al. (2013b)}&&&&&&&\\

CFHT-NBJ-1709 & 22:19:31.92 & +00:36:11.6        & 0.8133 & 21.3        &  2.1          &  0.42\,$\pm$\,0.06        & $10.7\pm0.1$                      &  8.5                      &    55.          & 0.5\,$\pm$\,0.10  \\
CFHT-NBJ-1713 & 22:19:21.34 & +00:36:42.7        & 0.7639 & 21.1        &  3.9        &  ...                                     & $10.0\pm0.2$                      &  7.4                      &         ...               & ...                 \\ 
CFHT-NBJ-1721 & 22:19:24.10 & +00:37:11.2        & 0.8144 & 20.0        &  5.1         &  0.62\,$\pm$\,0.06        & $10.8\pm0.1$                      & 13.9                       &    240.          & 0.6\,$\pm$\,0.2   \\
CFHT-NBJ-1724 & 22:19:27.27 & +00:37:31.3        & 0.8117 & 21.4        &  4.7        &  0.36\,$\pm$\,0.08        &  $10.1\pm0.1 $                    &  4.3                       &         ...                & ...                 \\
CFHT-NBJ-1733 & 22:19:43.57 & +00:38:22.1        & 0.7731 & 22.2        &  3.8      &  0.19\,$\pm$\,0.03        &  $9.7\pm0.3 $                   &  7.6                      &    90.          & 1.4\,$\pm$\,0.5  \\ 
CFHT-NBJ-1739 & 22:19:42.27 & +00:38:31.6        & 0.8042 & 20.1        &  6.0       &  0.40\,$\pm$\,0.05        & $10.6 \pm0.2  $                   &  11.4                       &    247.          & 0.5\,$\pm$\,0.1   \\
CFHT-NBJ-1740 & 22:19:18.60 & +00:38:43.9        & 0.8128 & 21.2        &  5.0         &  0.32\,$\pm$\,0.05        & $10.4 \pm0.1   $                  &  8.9                       &    217.        & 0.3\,$\pm$\,0.1   \\
CFHT-NBJ-1745 & 22:19:29.51 & +00:38:52.1        & 0.8174 & 22.0        &  4.1      &  0.16\,$\pm$\,0.02        &  $9.8  \pm0.3    $                & 5.6                       &    211.         & 0.2\,$\pm$\,0.1   \\
CFHT-NBJ-1759 & 22:19:41.42 & +00:39:25.4        & 0.8035 & 20.3        &  4.1         &  0.39\,$\pm$\,0.03        & $10.3\pm0.2    $                  &  12.9                       &    275.          & 0.2\,$\pm$\,0.1  \\
CFHT-NBJ-1770 & 22:19:27.66 & +00:40:14.3        & 0.7731 & 21.7        &  3.9        &  0.05\,$\pm$\,0.01        &  $9.9 \pm0.3    $                 &  10.4                       &    144.       & 0.4\,$\pm$\,0.2   \\ 
CFHT-NBJ-1774 & 22:19:30.59 & +00:40:31.5        & 0.8127 & 21.7        &  3.8         &  0.19\,$\pm$\,0.03        &  $9.8\pm0.2     $                 &  4.2                      &    50.         & 0.3\,$\pm$\,0.1   \\
CFHT-NBJ-1787 & 22:19:39.21 & +00:41:20.8        & 0.8132 & 20.5        &  6.5         &  0.41\,$\pm$\,0.04        & $10.6\pm0.2    $                  &  12.0                       &    255.          & 0.3\,$\pm$\,0.1   \\
CFHT-NBJ-1789 & 22:19:23.19 & +00:41:23.8        & 0.8130 & 20.6        &  9.5         &  0.32\,$\pm$\,0.02        & $10.6\pm0.1    $                  &  11.8                       &    253.          & 0.1\,$\pm$\,0.1   \\
CFHT-NBJ-1790 & 22:19:24.69 & +00:41:26.1        & 0.8124 & 22.0        &  1.7          &  0.30\,$\pm$\,0.05        & $9.9\pm0.3      $               &  4.7                       &    30.          & 0.4\,$\pm$\,0.2   \\
CFHT-NBJ-1793 & 22:19:30.60 & +00:41:35.1        & 0.8161 & 21.3        &  9.3           &  0.30\,$\pm$\,0.04        & $10.2\pm0.2$                      &  7.8                       &         ...                & ...                 \\
CFHT-NBJ-1795 & 22:19:32.44 & +00:41:42.3        & 0.8095 & 21.5        &  3.0           &  0.32\,$\pm$\,0.04        &  $9.8\pm0.2$                     &  6.5                       &    53.          & 0.5\,$\pm$\,0.1   \\
\hline
\end{tabular}
\end{center}
\end{table*}

\subsection{Metallicity Gradients}
\label{sec:metgrad}
We derive the metal content of our galaxies using emission line ratios. The gas phase abundance of Oxygen [$\rm 12+\log(O/H)$] for the sample can be estimated from the ratio of the [NII] to $\rm H\alpha$ lines \citep{Alloin79, denicolo2002,kewley2002}. This is often referred to as the N2 method, where

\begin{equation}
{\rm N2}=\log (f_{\rm [NII]}/f_{\rm H_\alpha})
\label{eq:n2}
\end{equation}

To convert from N2 to Oxygen abundance we use the conversion of \cite{Pettini2004}, which is appropriate for high redshift star-forming galaxies, where:

\begin{equation}
12+\log(\rm O/H)=8.9+0.57\rm N2
\label{eq:pet}
\end{equation}

We first derived metallicities within an aperture of diameter $1.2''$ for comparison with our Subaru FMOS study of HiZELS galaxy metallicities \citep{stott2013b}. The median metallicity of the sample is $12+\log(\rm O/H)=8.63\pm0.11$, consistent with the solar value of $8.66\pm0.05$ \citep{asplund2004}. In Table \ref{tab:samp} we show the uncertainty in the [NII] to $\rm H\alpha$ ratio derived from the errors of the line profile fitting, however we note that the \cite{Pettini2004} metallicity calibration has a $1\sigma$ scatter of 0.18\,dex. The mass metallicity relation for the KMOS-HiZELS galaxies is in agreement with both the low redshift SDSS study of \cite{tremonti2004} and the $z\sim0.8-1.5$ relation displayed in \cite{stott2013b} and is therefore consistent with no chemical abundance evolution since $z\sim1$ (this is discussed in \citealt{stott2013b}). In terms of AGN contamination only one of the galaxies in the entire sample has $\rm N2>0.0$ (CFHT-NBJ-C339) which may indicate that it is an AGN \citep{kew2001}.

\begin{figure*}
   \centering
\includegraphics[scale=0.3, trim=0 15 30 0, clip=true]{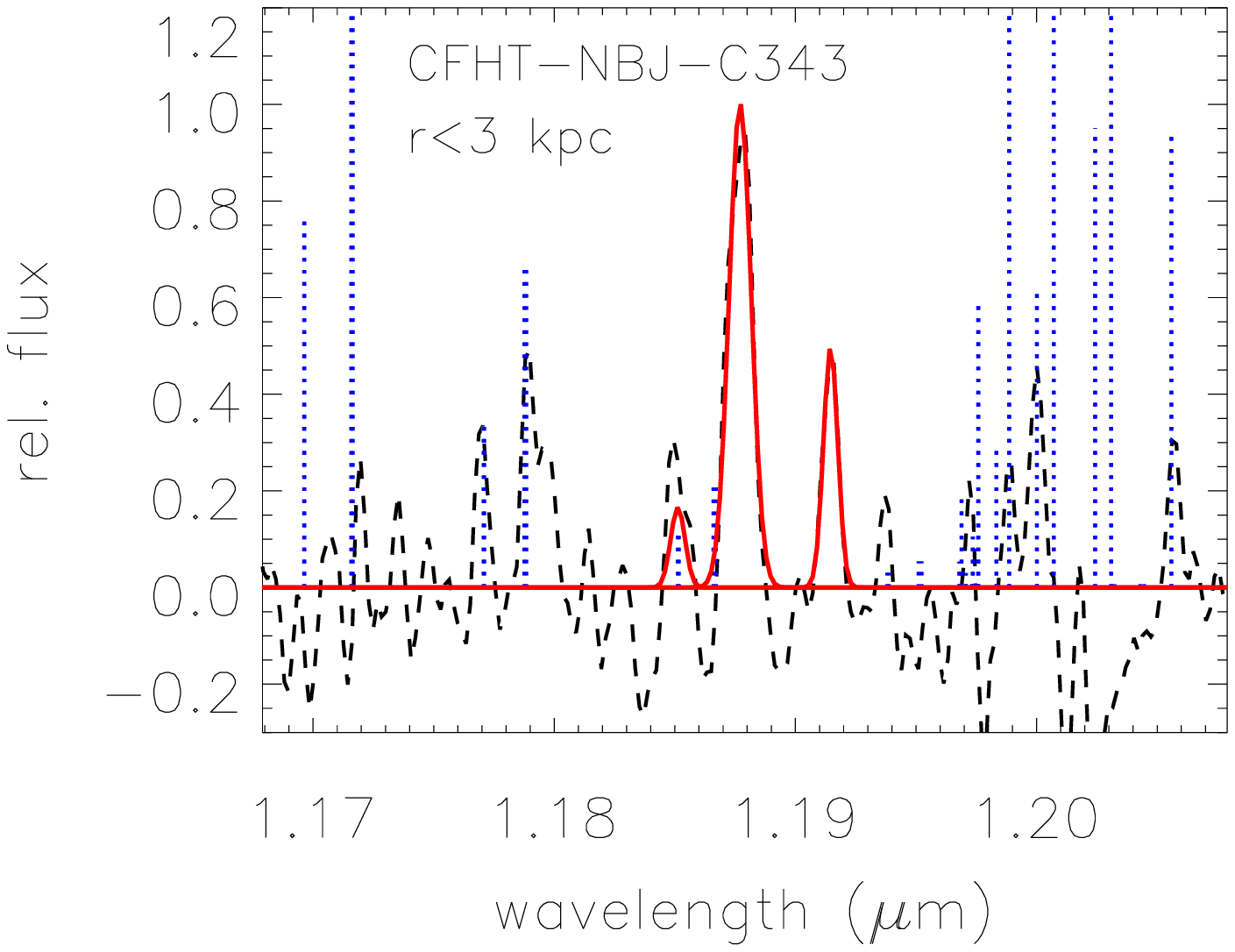} 
\includegraphics[scale=0.3, trim=115 15 30 0, clip=true]{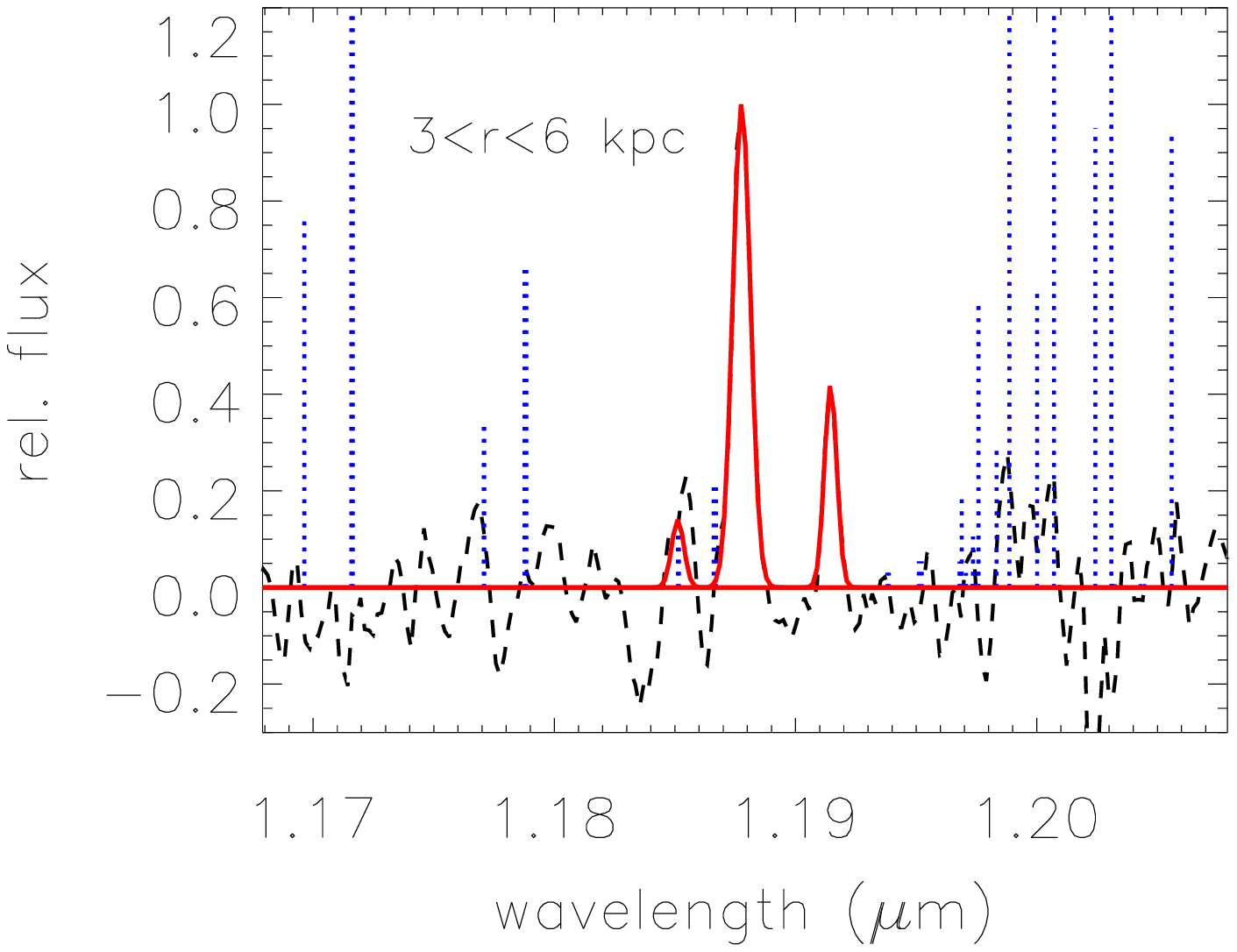} 
\includegraphics[scale=0.3, trim=115 15 0 0, clip=true]{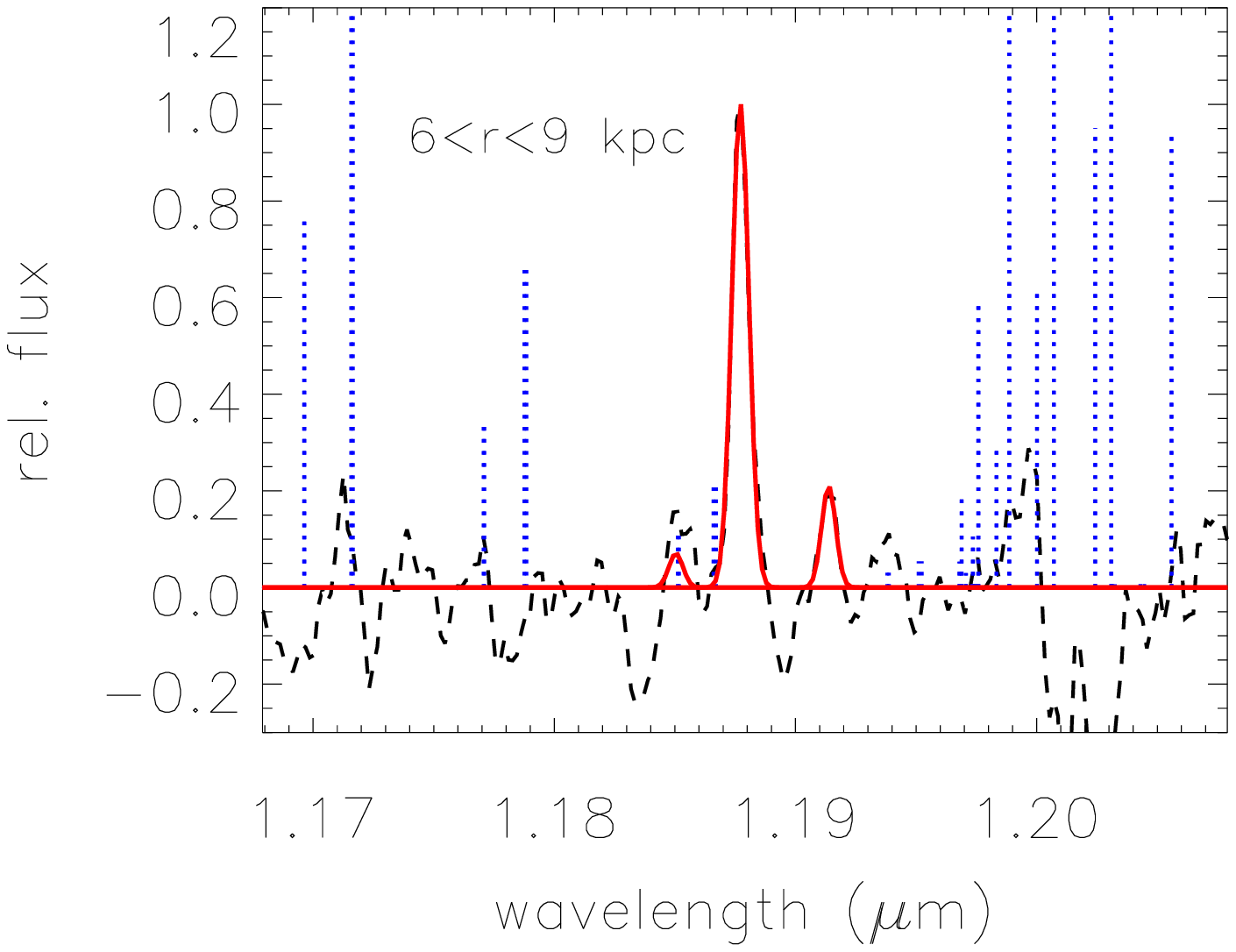} 
\includegraphics[scale=0.3, trim=0 15 0 0, clip=true]{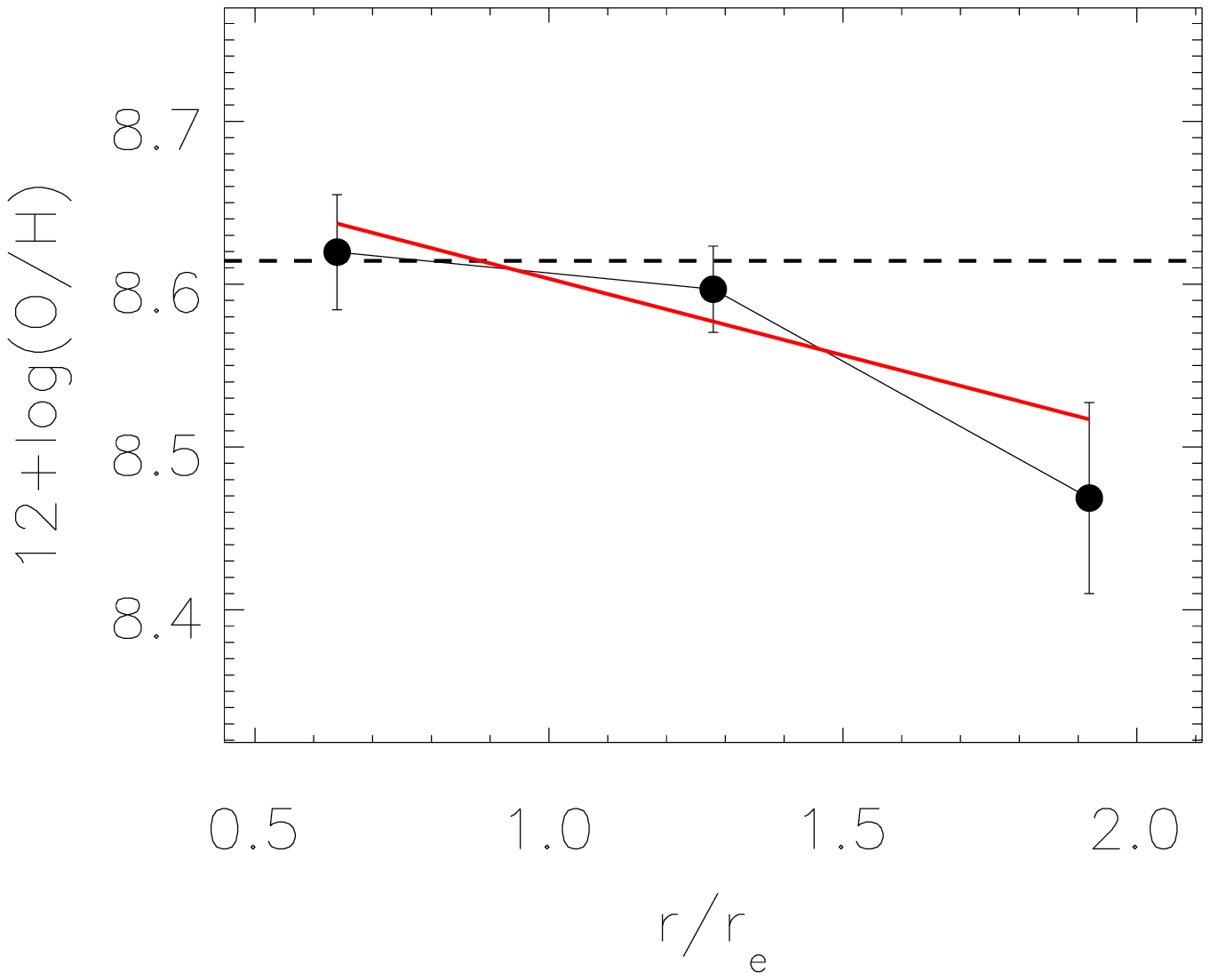}

\includegraphics[scale=0.3, trim=0 15 30 0, clip=true]{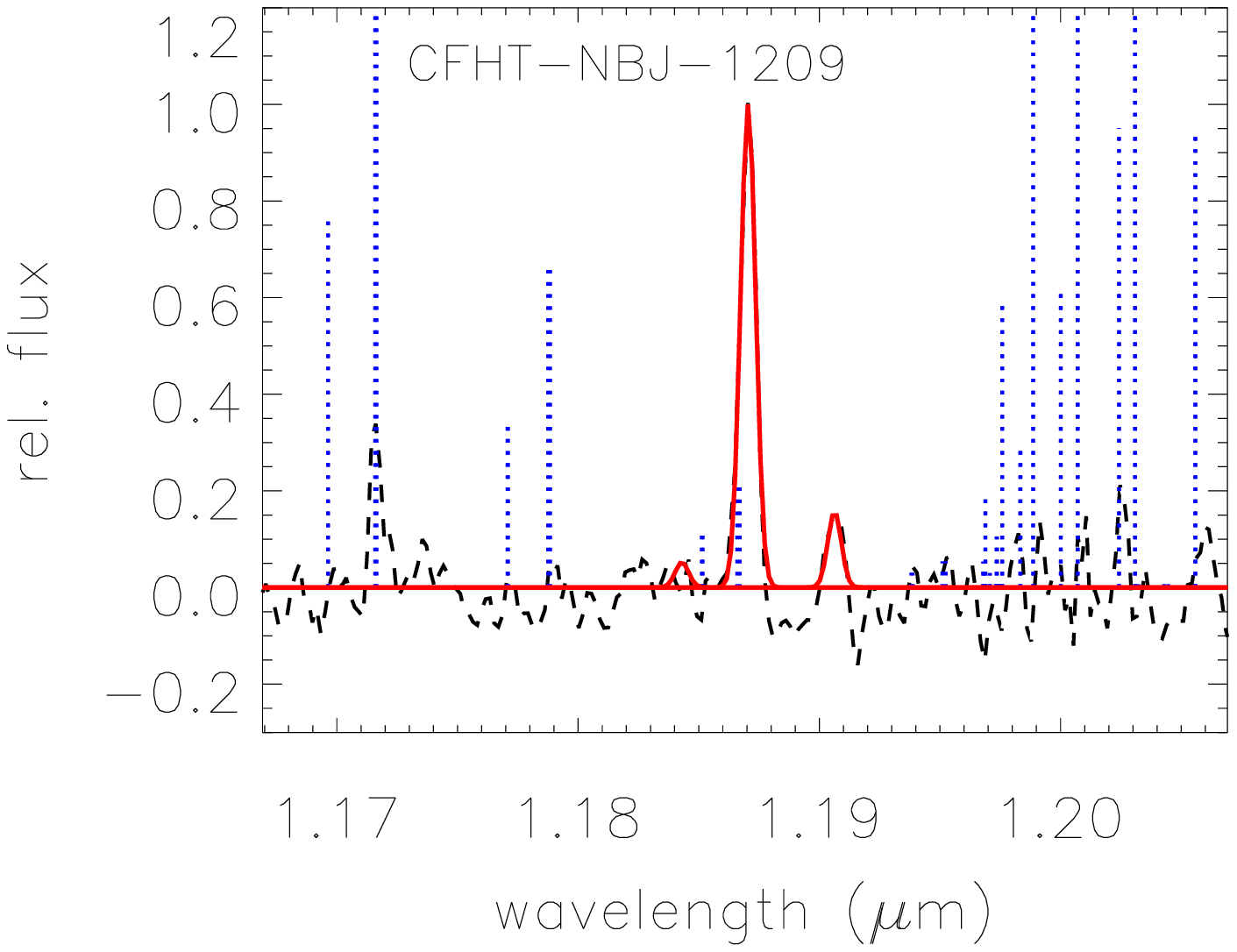} 
\includegraphics[scale=0.3, trim=115  15 30 0, clip=true]{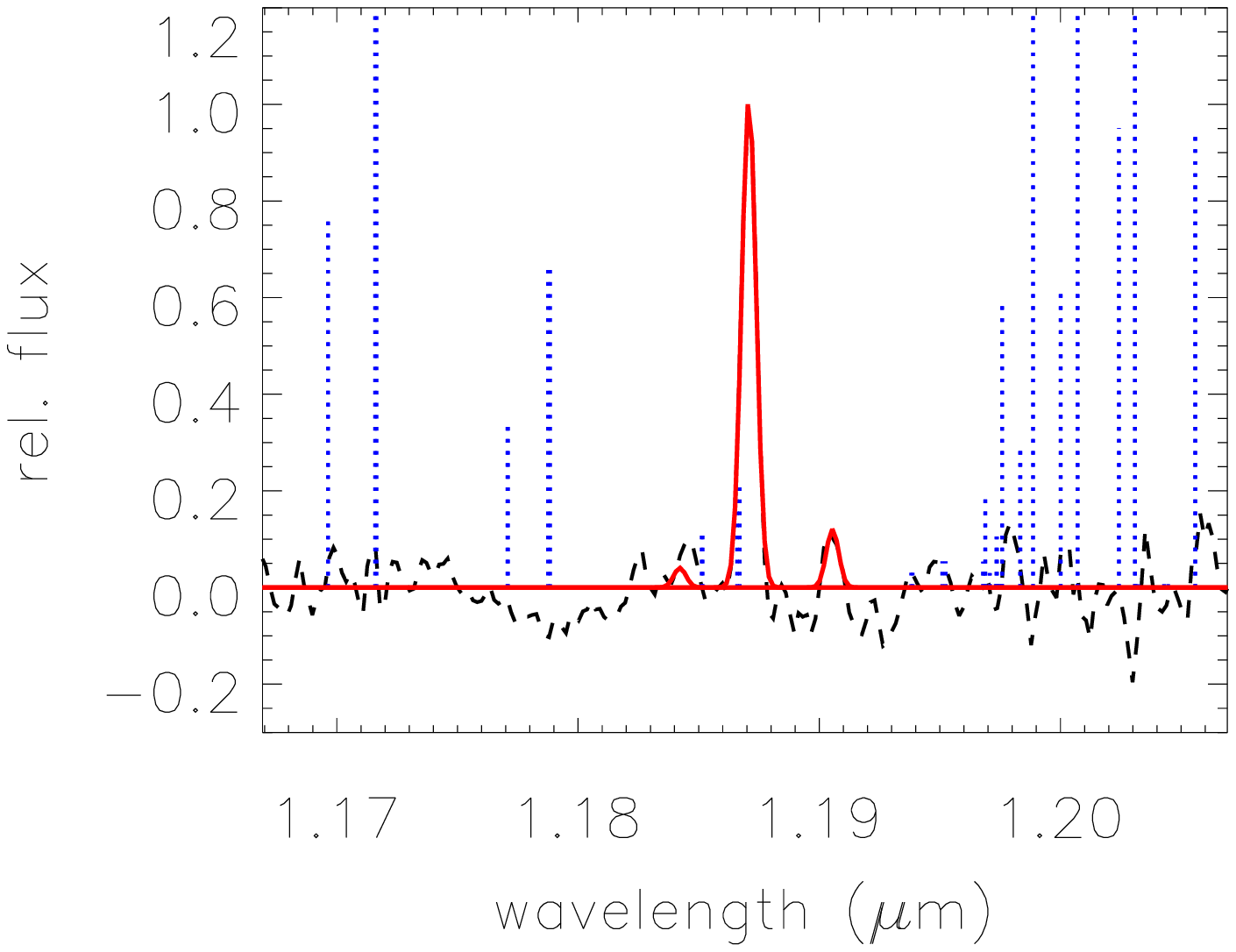} 
\includegraphics[scale=0.3, trim=115 15 0 0, clip=true]{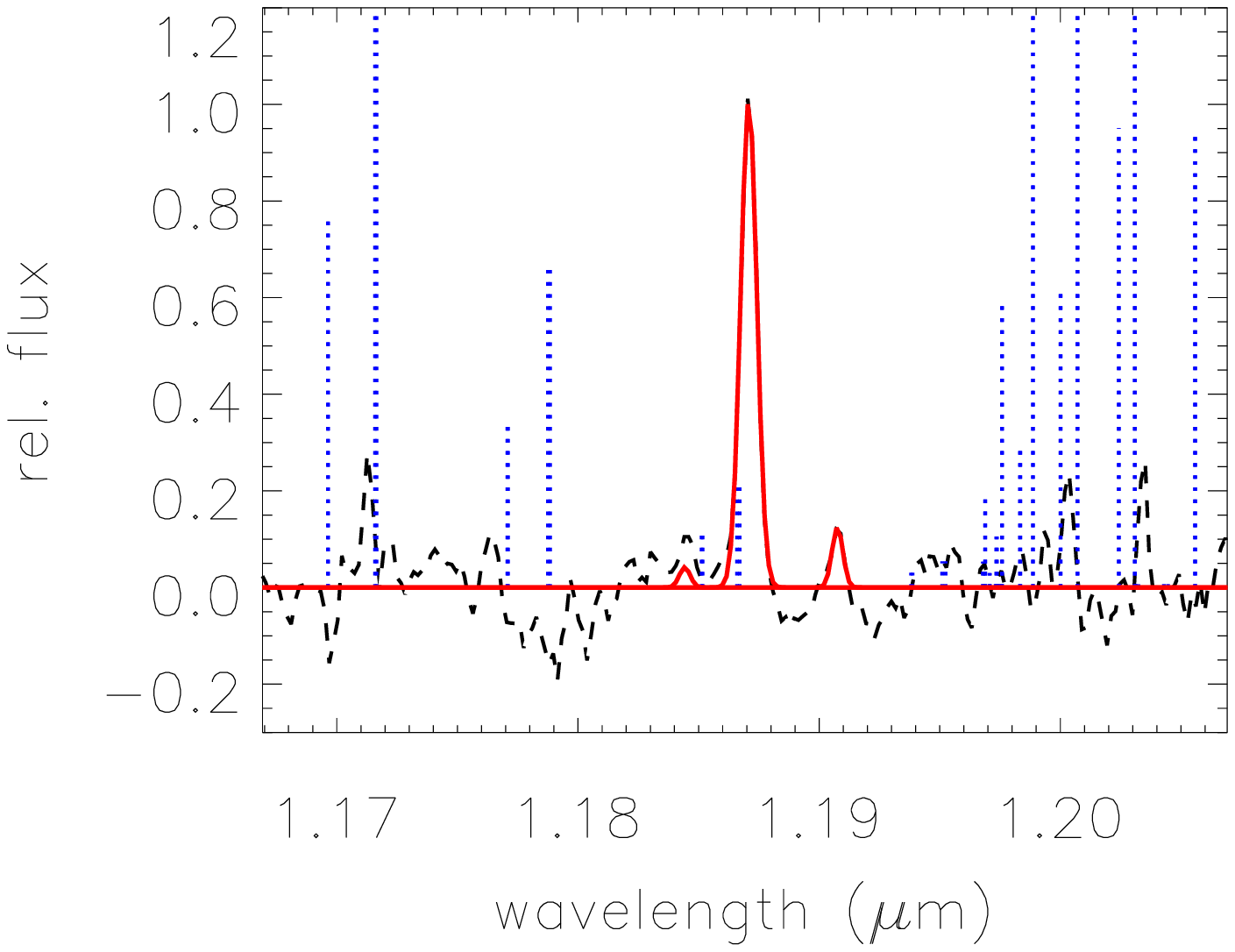} 
\includegraphics[scale=0.3, trim=0 15 0 0, clip=true]{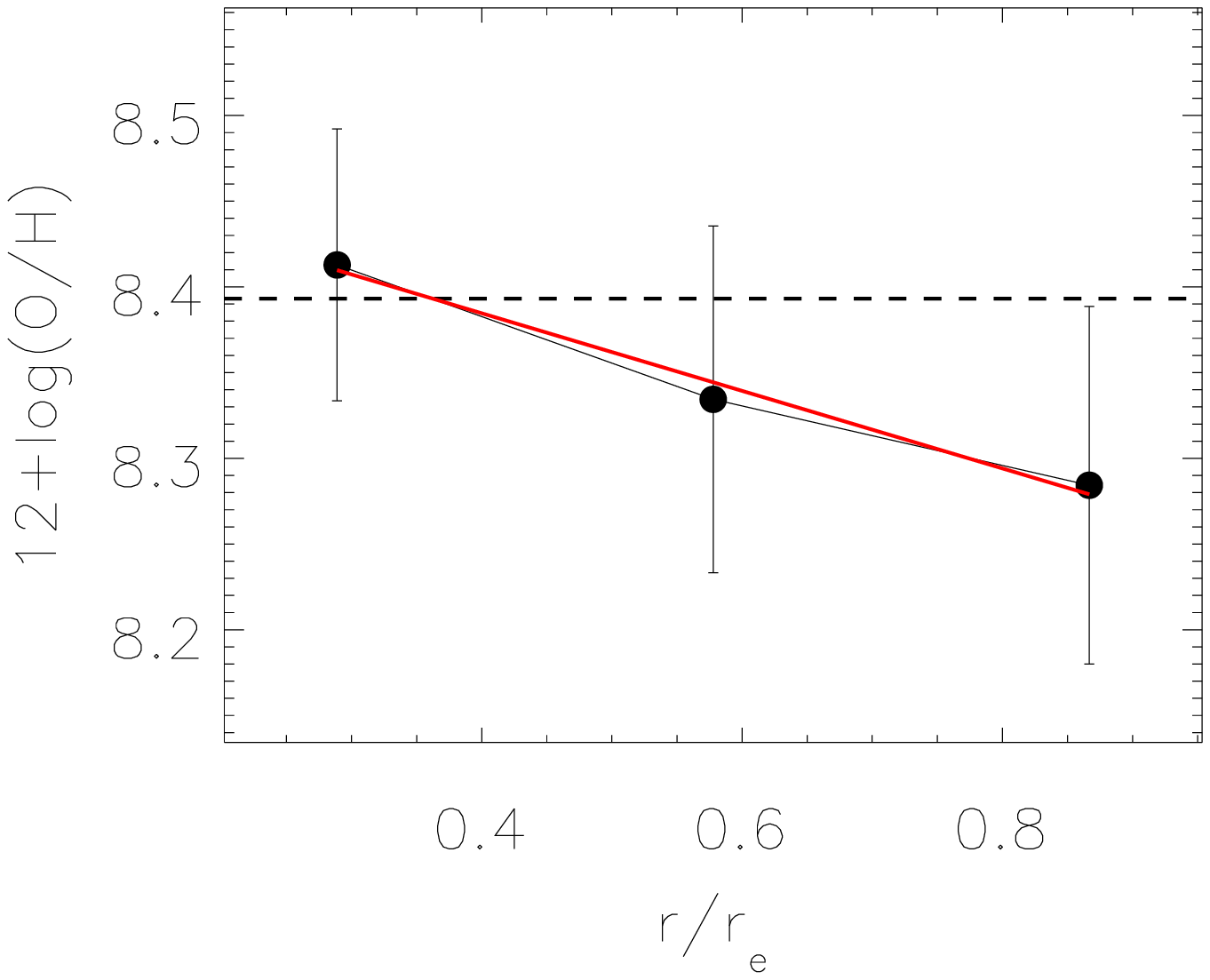} 

\includegraphics[scale=0.3, trim=0 15 30 30, clip=true]{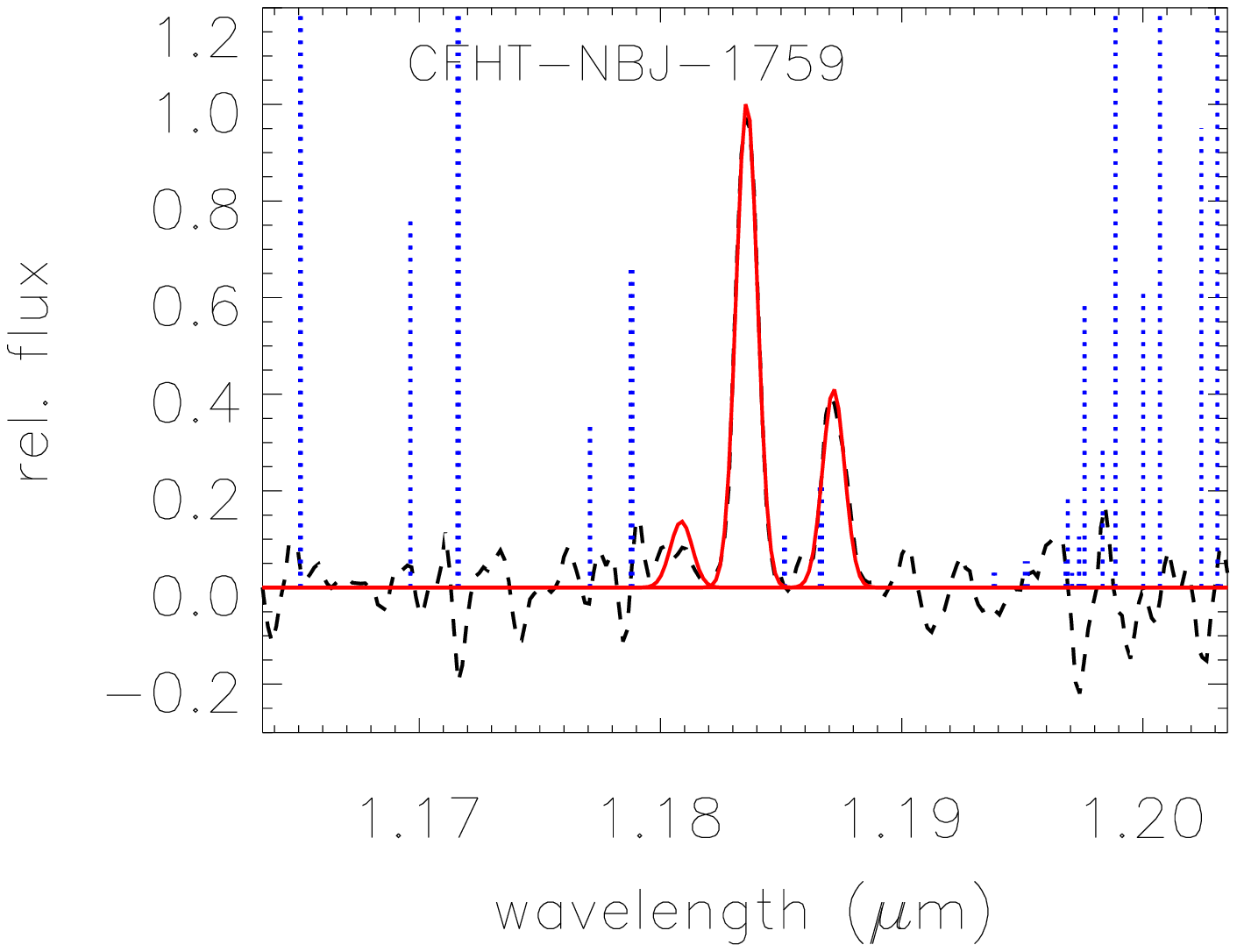} 
\includegraphics[scale=0.3, trim=115 15 30 0, clip=true]{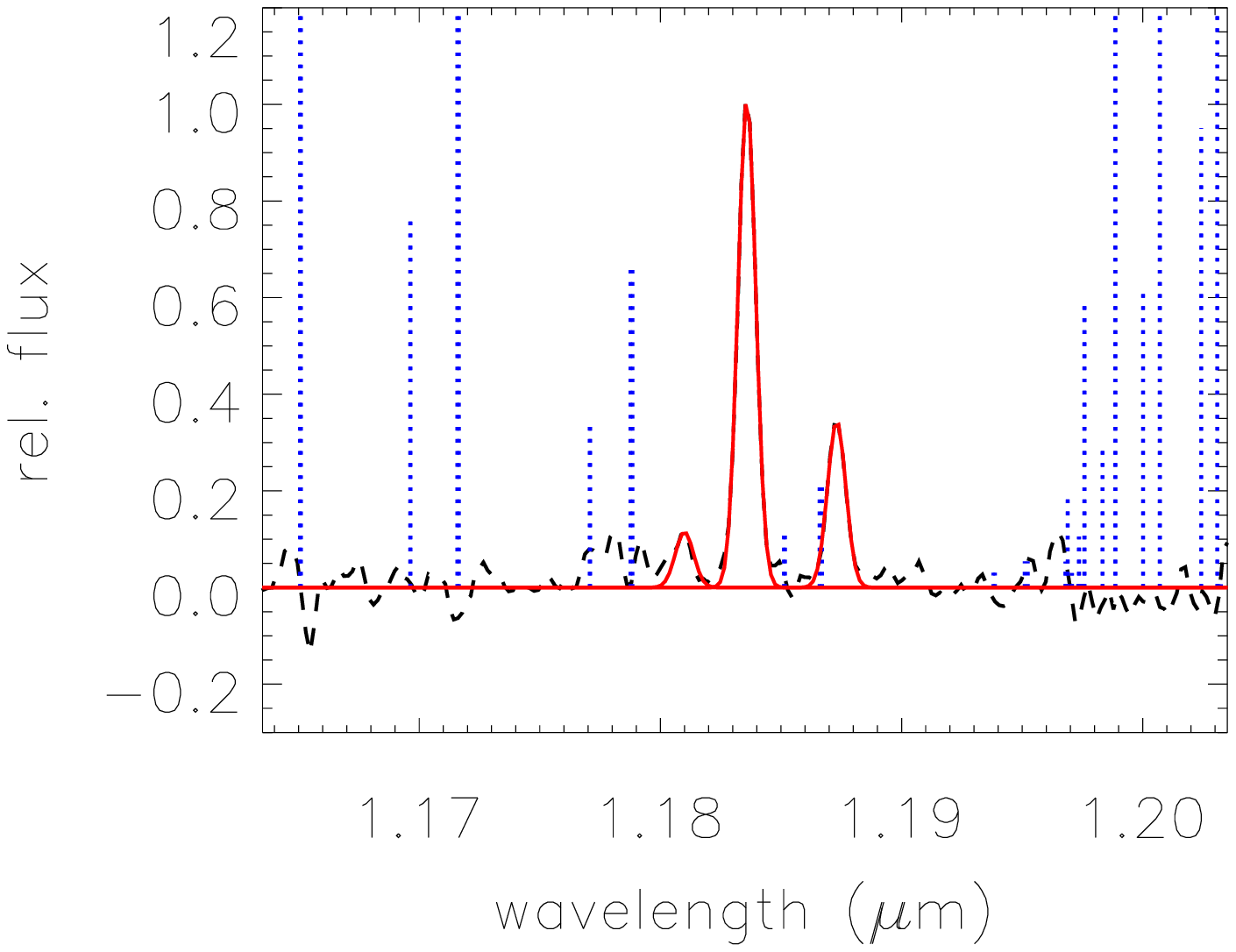} 
\includegraphics[scale=0.3, trim=115 15 0 0, clip=true]{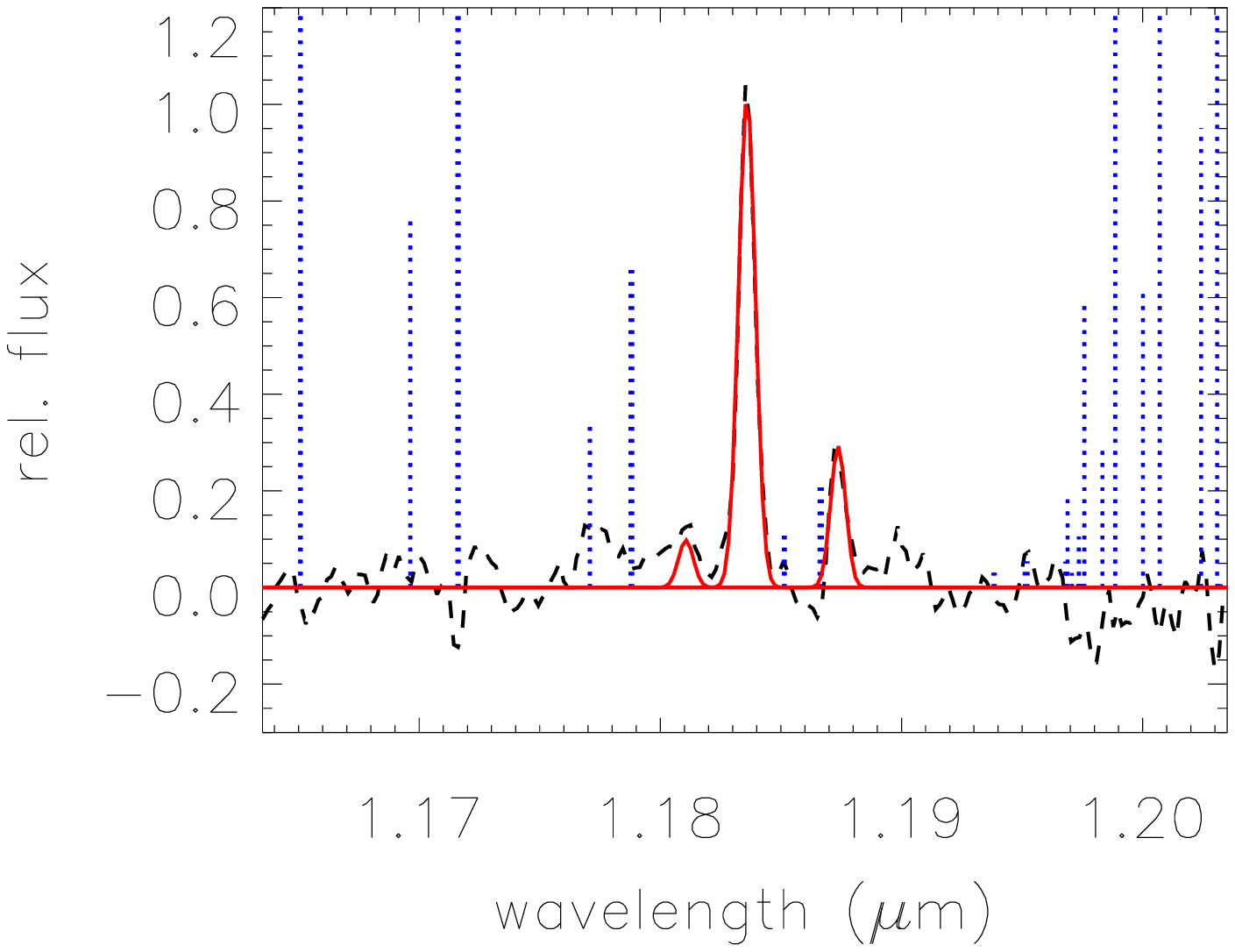} 
\includegraphics[scale=0.3, trim=0 15 0 0, clip=true]{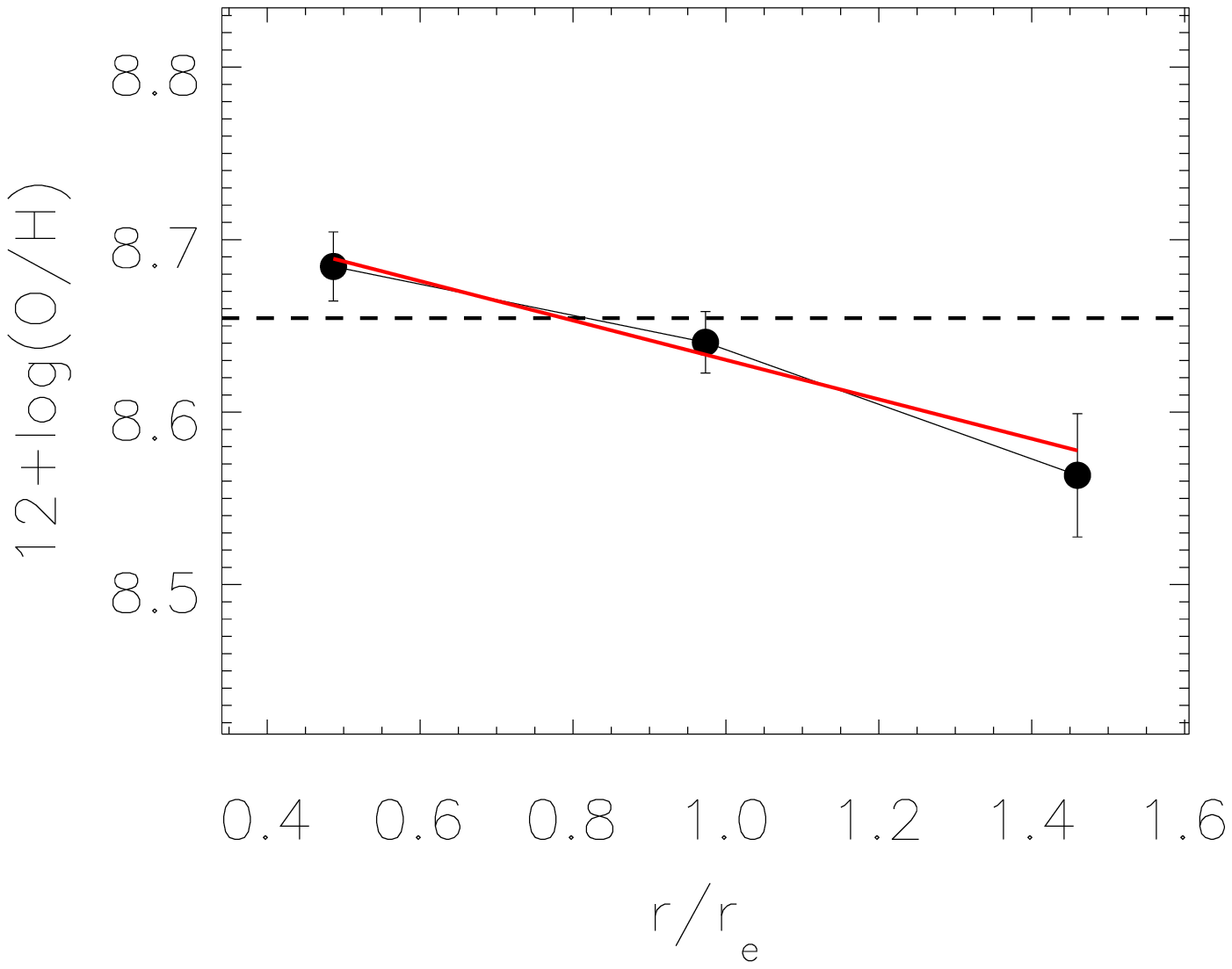}

\includegraphics[scale=0.3, trim=0 15 30 0, clip=true]{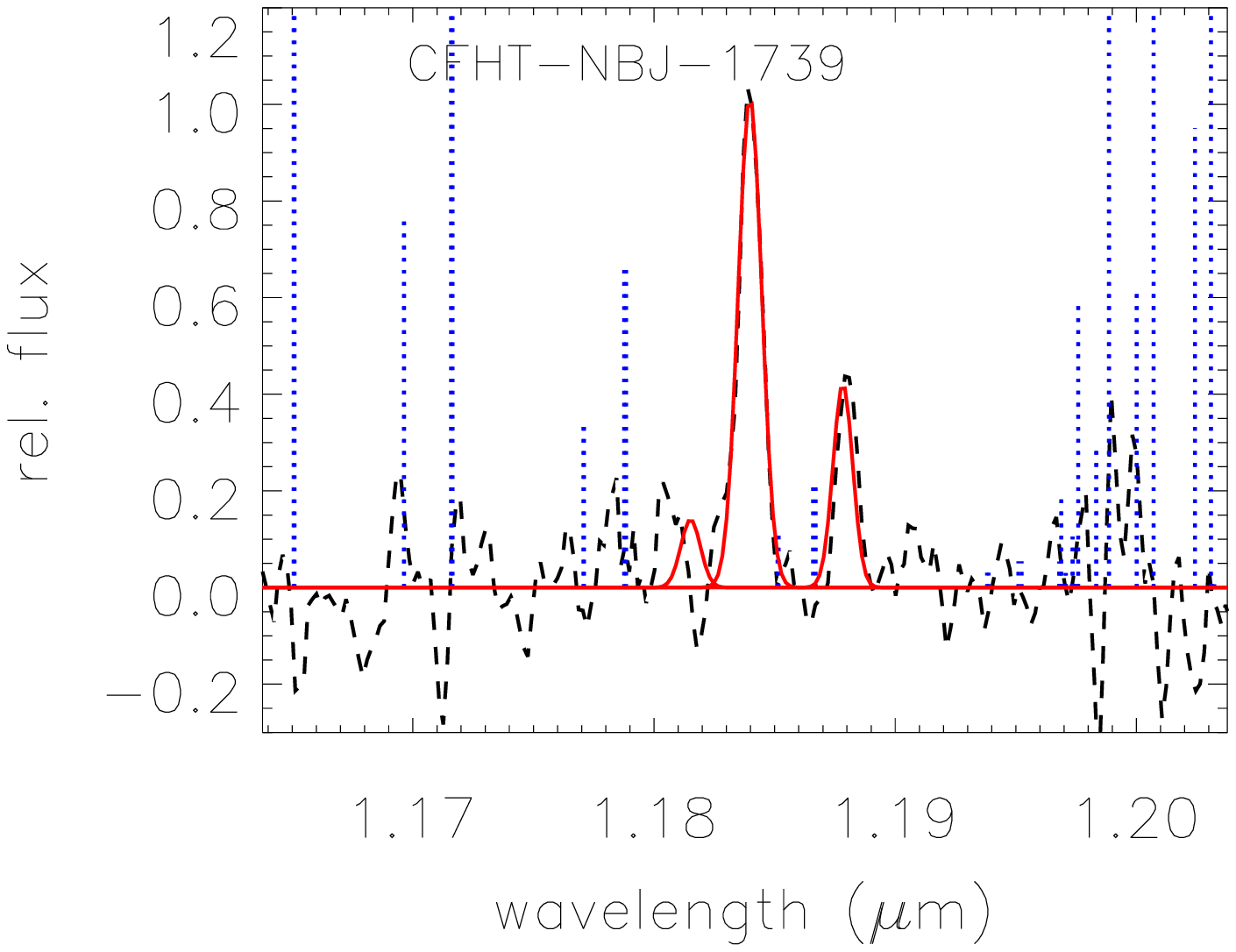} 
\includegraphics[scale=0.3, trim=115 15 30 0, clip=true]{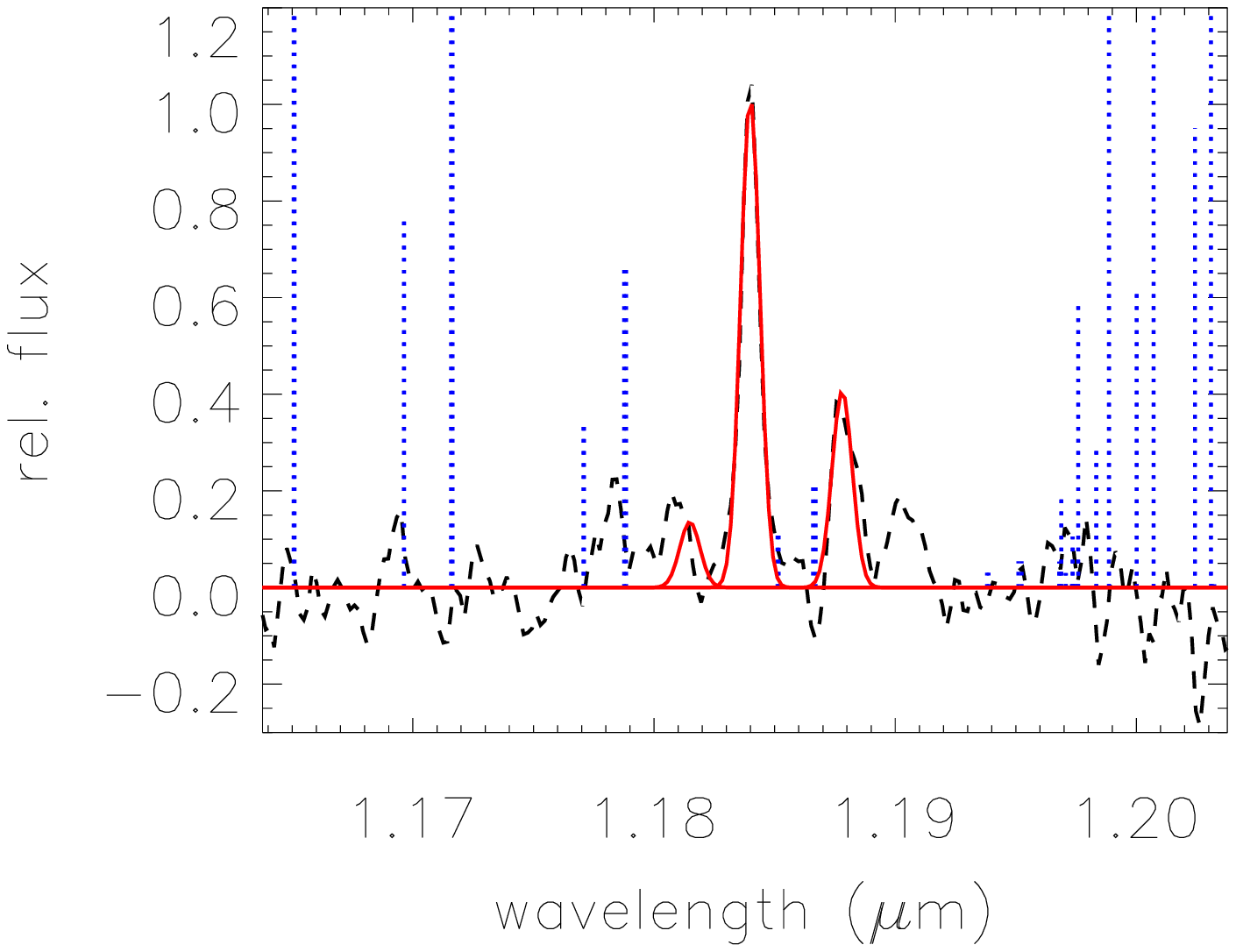} 
\includegraphics[scale=0.3, trim=115 15 0 0, clip=true]{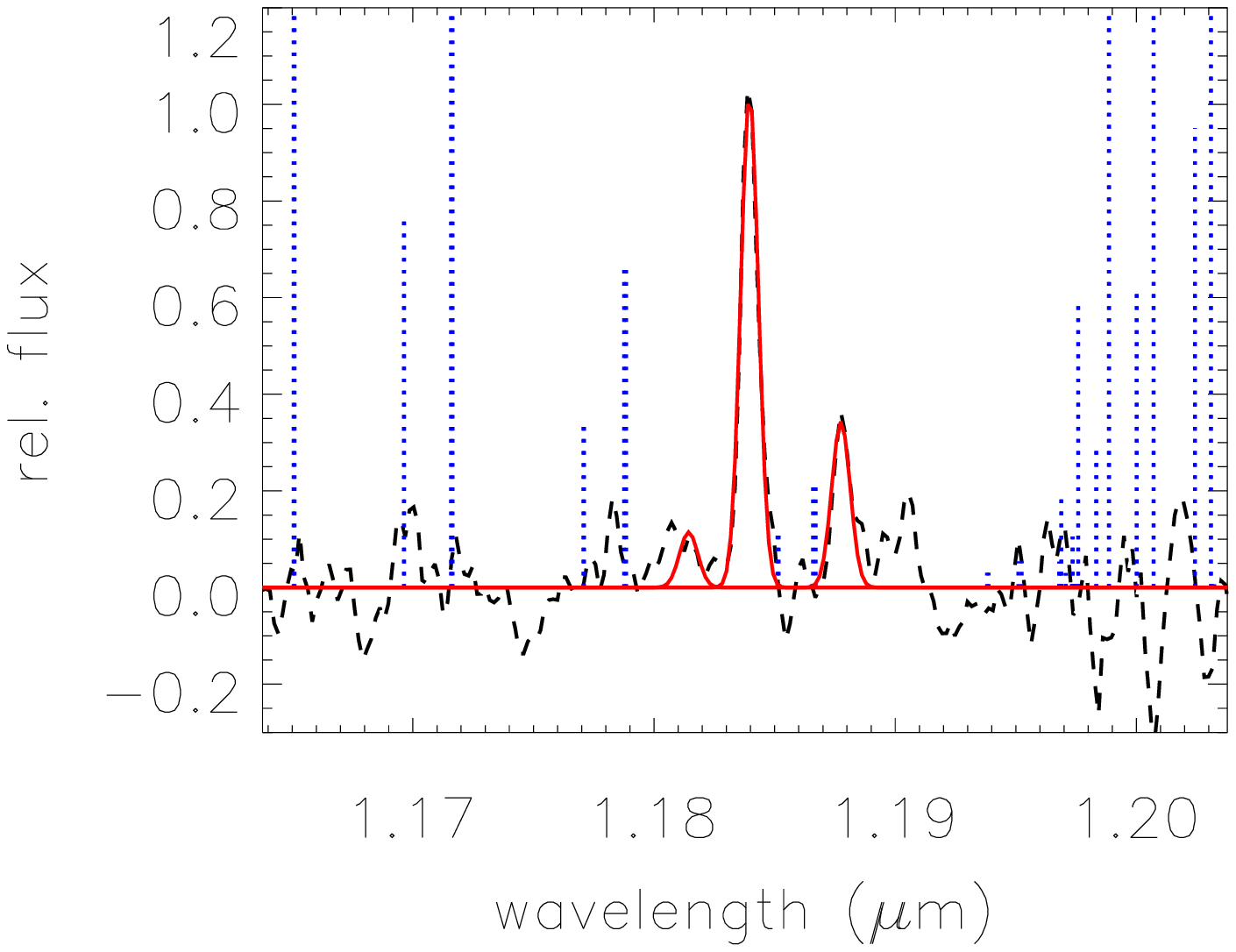} 
\includegraphics[scale=0.3, trim=0 15 0 0, clip=true]{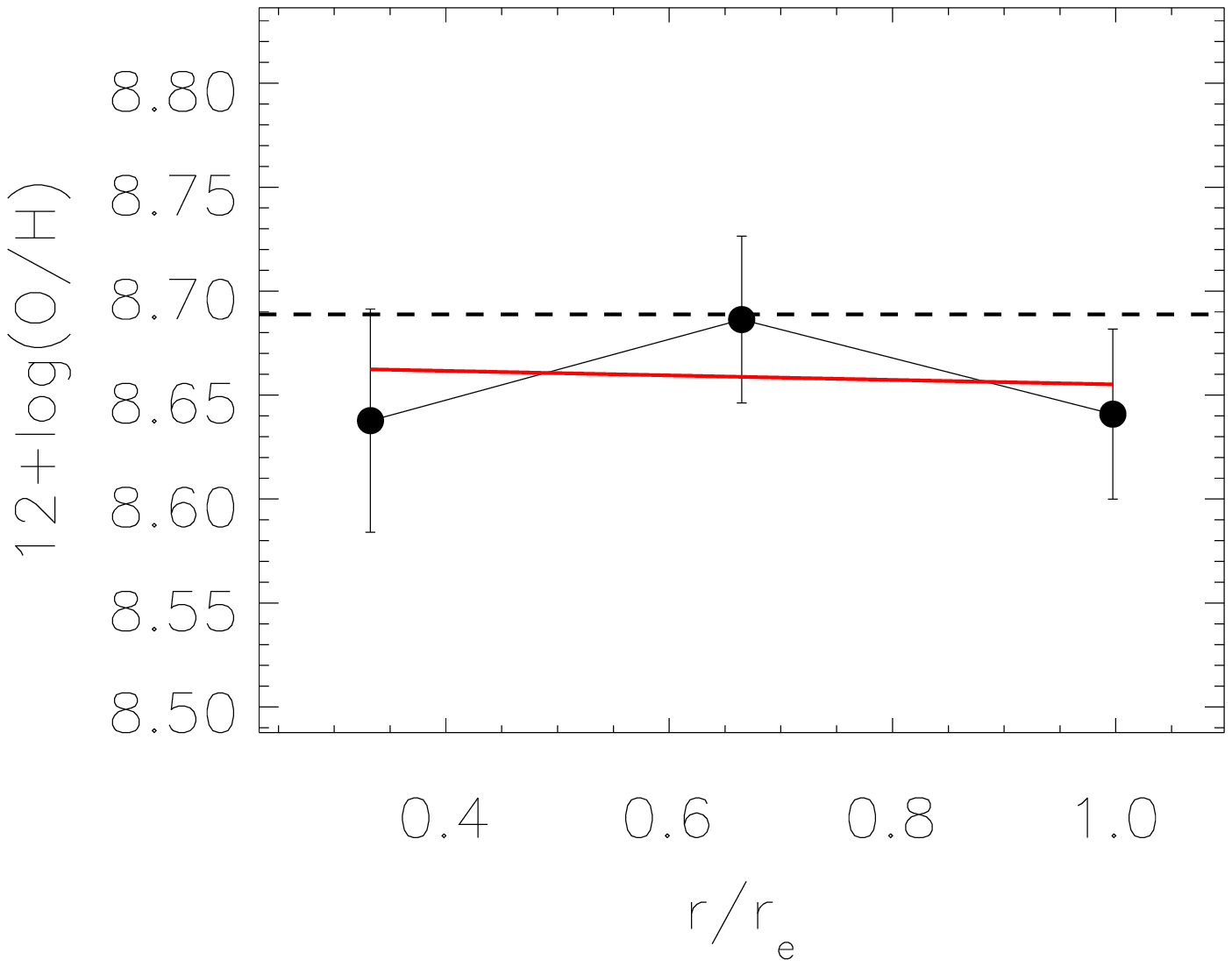}

\includegraphics[scale=0.3, trim=0 15 30 0, clip=true]{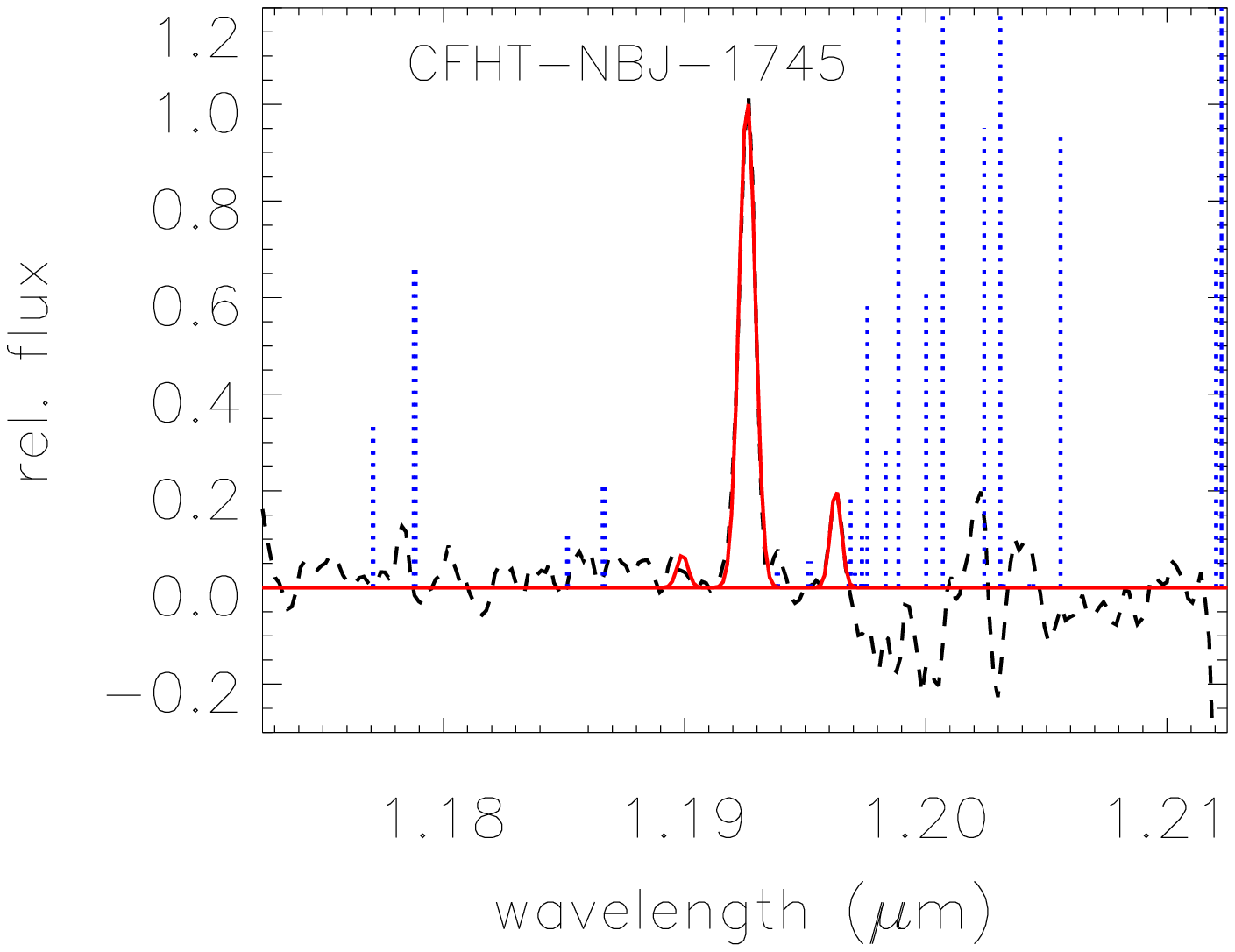} 
\includegraphics[scale=0.3, trim=115 15 30 0, clip=true]{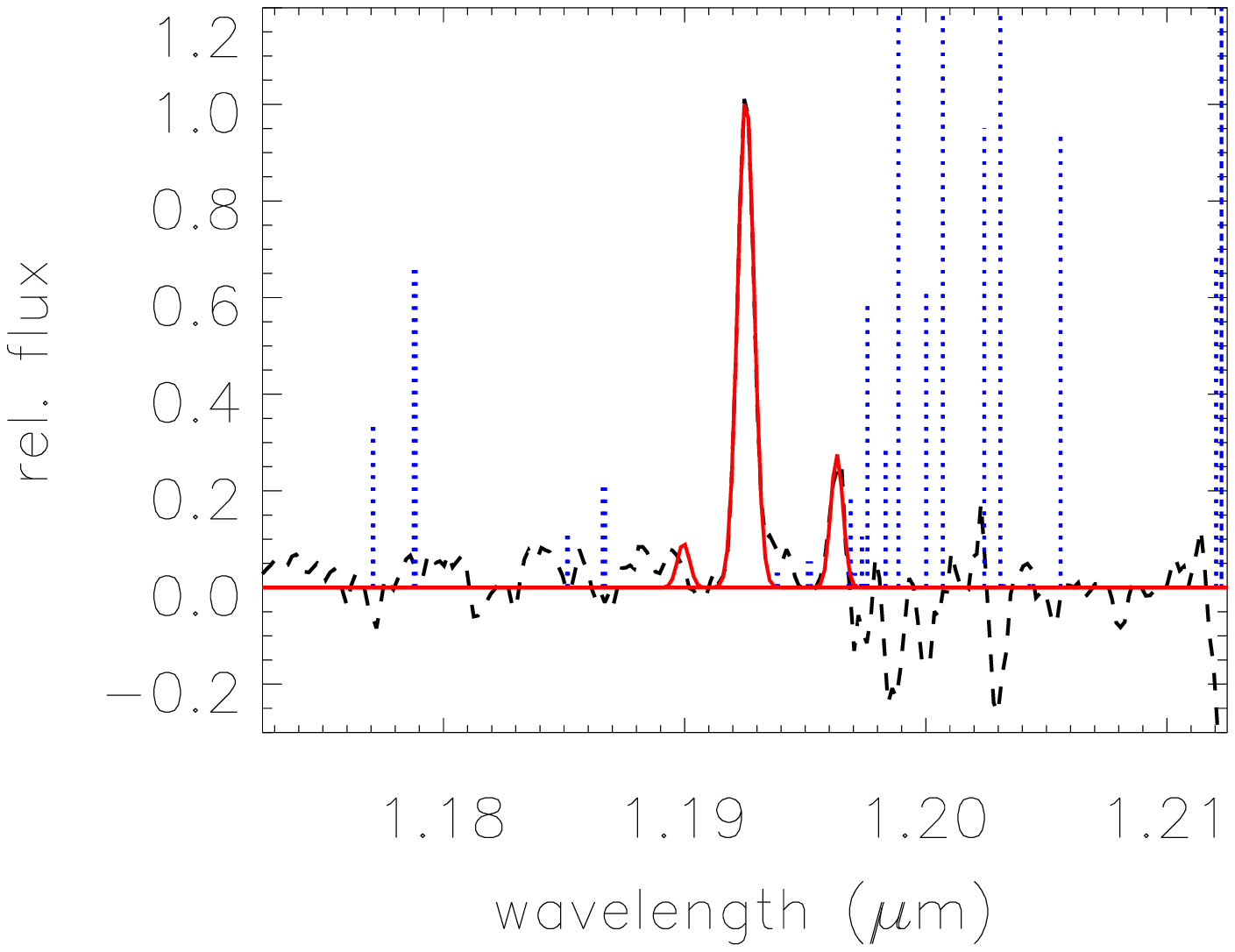} 
\includegraphics[scale=0.3, trim=115 15 0 0, clip=true]{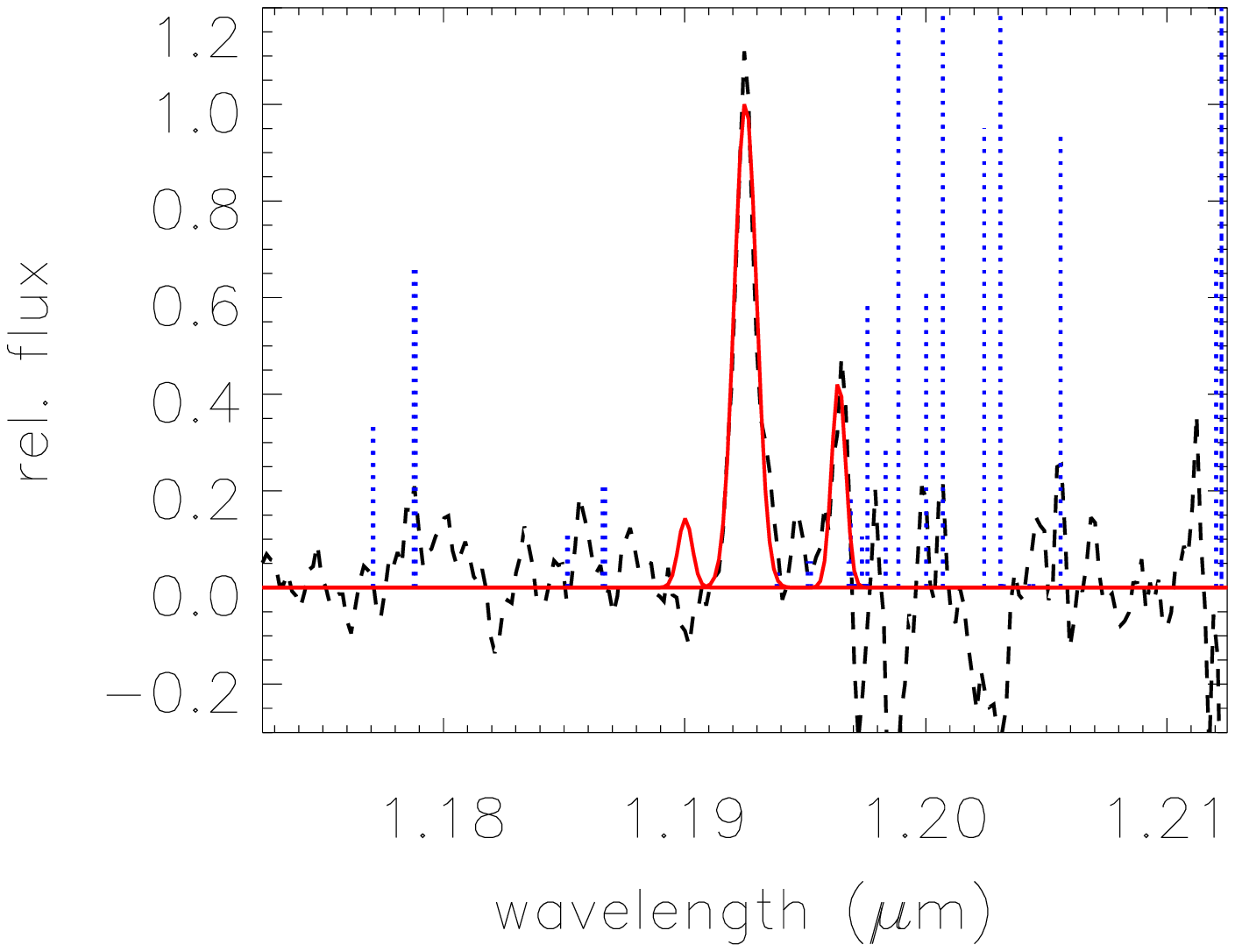} 
\includegraphics[scale=0.3, trim=0 15 0 0, clip=true]{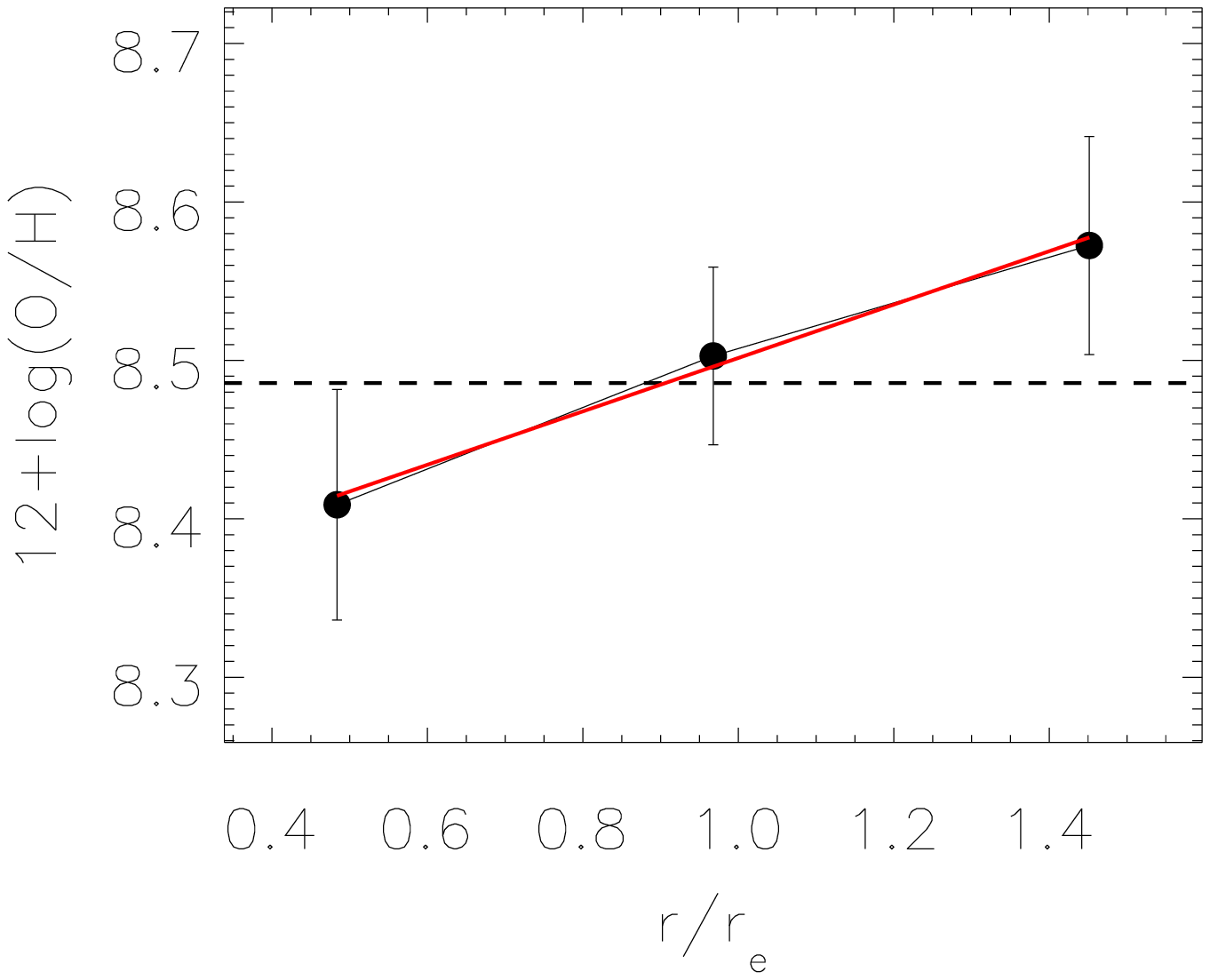}

\caption[]{The metallicity gradients for five galaxies from the KMOS-HiZELS sample. \emph{Left:} These are the individual 1-dimensional spectra from three concentric annuli at increasing galactocentric radius ($r<3$, $3<r<6$, $6<r<9$\,kpc). The red lines are fits to the $\rm H\alpha$ and [NII] emission lines. The dotted blue lines represent the location and relative flux of the sky emission. \emph{Right:} The metallicity derived from the ratio of $\rm [NII]/H\alpha$ plotted against galactocentric radius. The red line is a fit to the data points. The horizontal dashed line represents global metallicity value for the galaxy measured in a $1.2''$ diameter aperture.}
   \label{fig:met1}
\end{figure*}

To derive the metallicity gradients of the galaxies in our sample we extract the average metallicity within elliptical annuli at increasing galactocentric radii. The ellipticity of these annuli is derived from the inclination angle of the best fitting dynamical disc model, found in \S\ref{sec:dyn}. The typical seeing for the observations is $0.7''$ which corresponds to $\sim \rm 5 kpc$. Given this, we choose to measure the metallicities in galactocentric annuli encompassing the radii: $\rm <3$, $\rm 3-6$ and $\rm 6-9\,kpc$ (a discussion of the effects of seeing and inclination angle is provided in \S\ref{sec:see}). In order to do this we first subtract the velocity field of the best fitting dynamical disc model, found in \S\ref{sec:dyn}, from the data cube so that the $\rm H\alpha$ and [NII] emission lines are not broadened or superimposed. We then sum the IFU spectra in each of these annuli and fit the $\rm H{\alpha}$\,6563\AA\, and [NII]\,6583\AA\, emission lines in the resulting 1-D spectra with single Gaussian profiles in order to extract their total flux. For a detection we enforce $5 \sigma$ and $2 \sigma$ detection thresholds over the continuum level for $\rm H{\alpha}$ and [NII] respectively (following \citealt{stott2013b}). Examples of the spectra in each annulus for five galaxies from our sample are displayed in Fig. \ref{fig:met1}. To calculate the metallicity gradient we use a $\chi^{2}$ minimisation to fit a straight line to the metallicity as a function of galactocentric radius and present the gradient values in Table \ref{tab:massmet}. The metallicity gradient fits are also displayed in Fig. \ref{fig:met1} with the radius normalised to the effective radius of the galaxy for ease of comparison. In total we were able to extract metallicity gradients for 20 of the KMOS-HiZELS galaxies as the remainder had integrated [NII] lines which were either too low signal-to-noise or affected by the sky emission spectra. {The measured metallicity gradient values are robust to the inclusion of the error arising from the 0.18\,dex scatter in the \cite{Pettini2004} metallicity calibration, although this would increase the typical gradient errors quoted in Table \ref{tab:massmet} by a factor of $\gtrsim2$. }

There is no evidence for the central annuli of any galaxy being dominated by AGN contamination except for the potential AGN CFHT-NBJ-C339 identified above. This galaxy has a high central N2 value of 0.1, although we note it has a line ratio gradient consistent with being flat so there is no central concentration. However, the presence of unaccounted-for AGN may act to boost the central N2 values of our galaxies, with both \cite{wright2010} and \cite{Newman2013} finding that at $z>1$ the region of the BPT (Baldwin, Phillips \& Terlevich) diagram \citep{baldwin1981} at the boundary between star-forming galaxies and AGN contains some composite systems with spatially concentrated AGN imbedded within a star-forming galaxy. We do not have the $\rm H{\beta}$ and [OIII] line diagnostics available to perform a BPT diagram. However, if we assume that some fraction of our sample may be affected by hidden AGN then this would act to raise their central metallicities, steepening the negative metallicity gradients but flattening the positive ones. {A further source of uncertainty is shock excited gas due to winds, which could increase the N2 values at large galactic radii, acting on the measured metallicity gradients in the opposite sense to AGN contamination \citep{rich2010}.}

The average value of the metallicity gradient for our sample is $\rm \frac{\Delta Z}{\Delta \it{r}}=-0.002\pm0.007\,dex\,kpc^{-1}$. There are seven galaxies with a $>2\sigma$ significance of having a non-zero metallicity gradient with five of these having negative gradients and two positive. 

\begin{figure*}
   \centering
\includegraphics[scale=0.5, trim=0 0 20 0, clip=true]{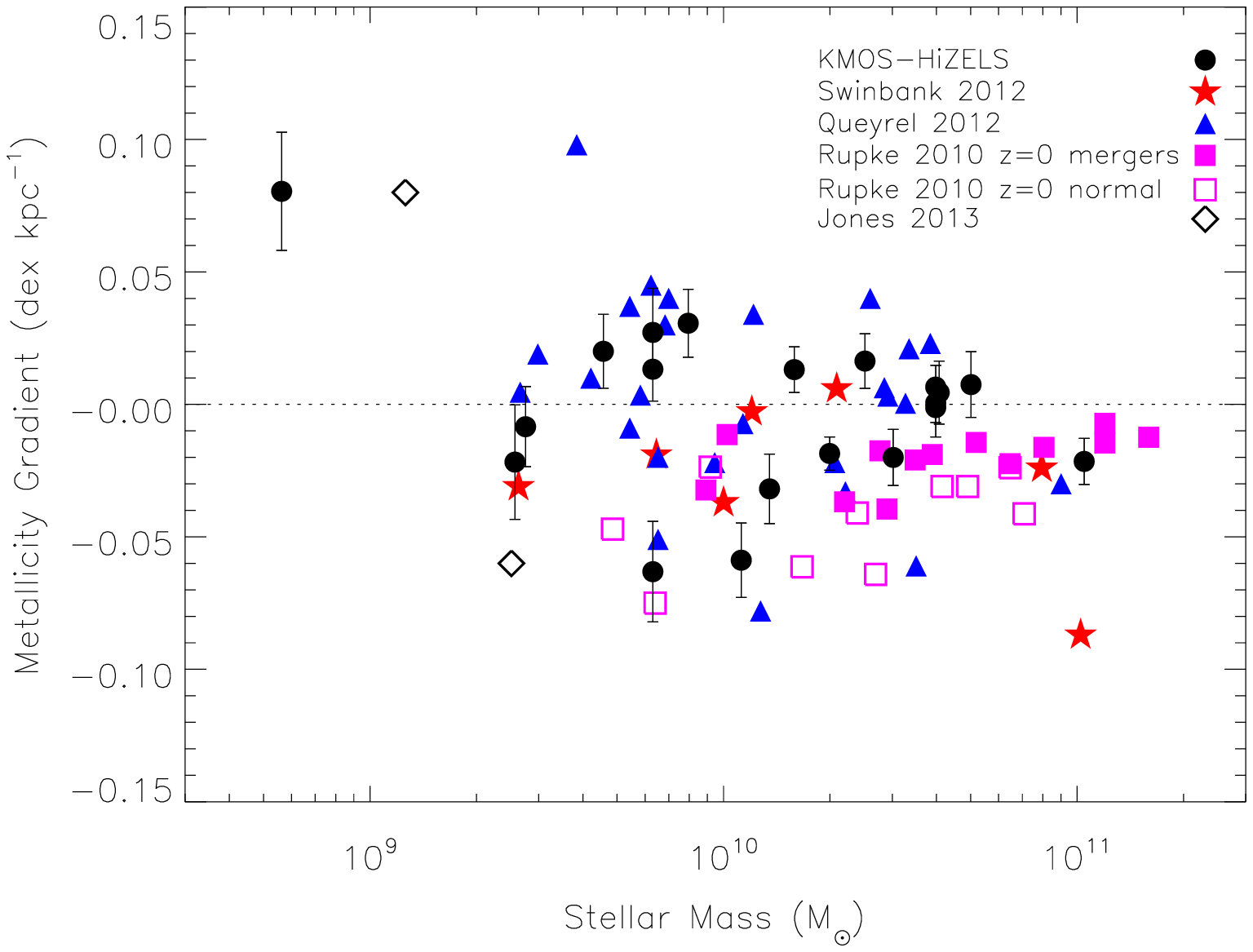} \includegraphics[scale=0.5, trim=70 0 0 0, clip=true]{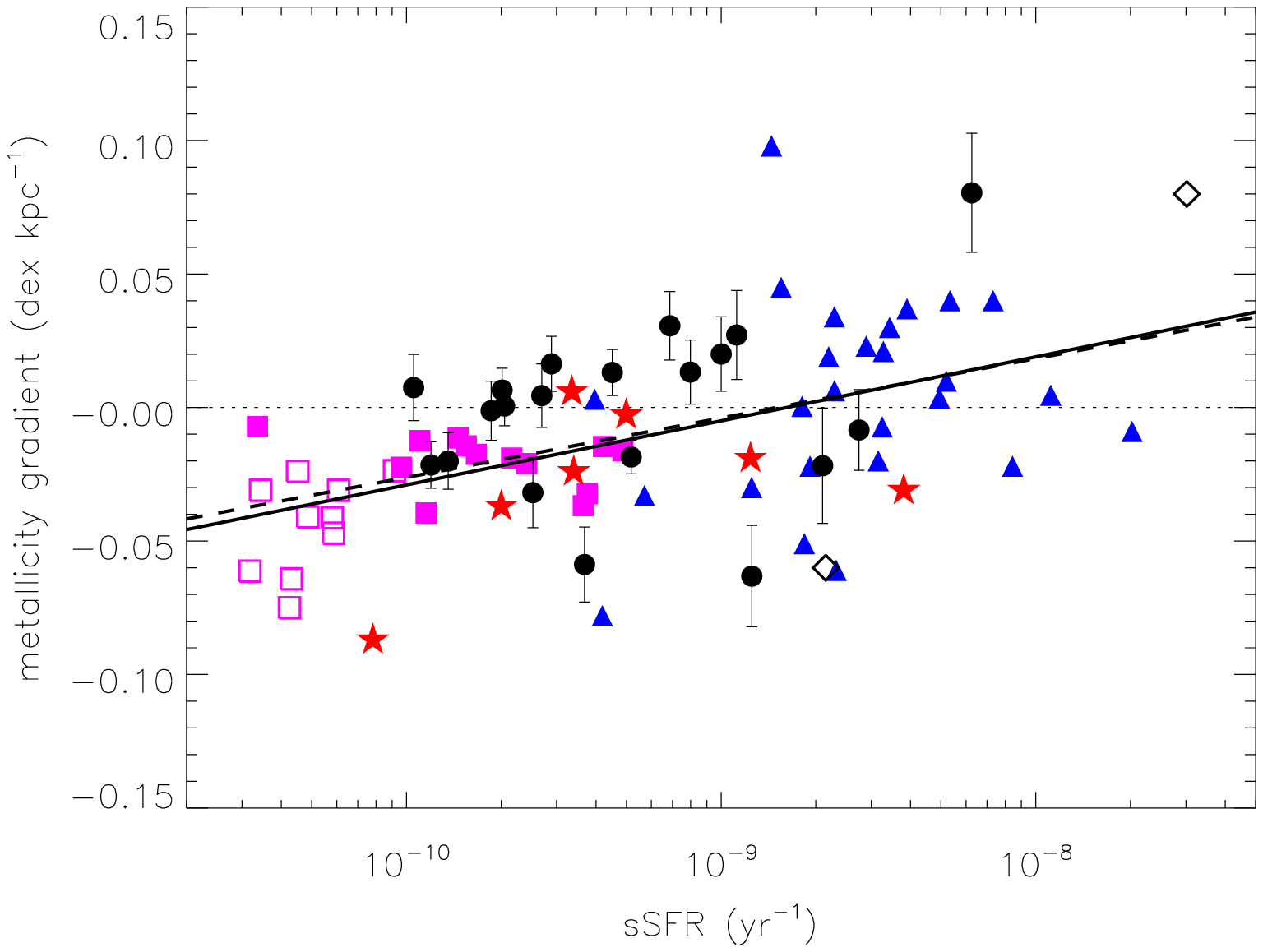}
\caption[]{ \emph{Left:} The metallicity gradient plotted against stellar mass for the KMOS-HiZELS sample. We include the high redshift samples of  \cite{swinbank2012} and \cite{queyrel2012} and the local samples of normal and merging systems from \cite{rupke2010b}. We also plot two galaxies from the lensing sample of \cite{jones2013}. \emph{Right:} The metallicity gradient plotted against sSFR the same samples. The solid line is a fit to the combined data of KMOS-HiZELS,  \cite{swinbank2012, queyrel2012} and \cite{rupke2010b}, which demonstrates that galaxies with higher sSFR tend to have more positive metallicity gradients. The dashed line is a fit to the high redshift galaxies only i.e. without the \cite{rupke2010b} samples.} 
   \label{fig:metgrad}
\end{figure*}

We look for correlations between the metallicity gradient and the global properties of the galaxies. In Fig. \ref{fig:metgrad} we plot the metallicity gradient against stellar mass and also include $z\sim1-2$ data points from the literature \citep{swinbank2012,queyrel2012,jones2013}, who all use the \cite{Pettini2004} N2 method to determine their metallicities. The stellar masses of the literature data are all estimated with a \cite{chabrier2003} IMF (as are the KMOS-HiZELS masses) except for \citep{queyrel2012}, which for consistency we correct from a \cite{salpeter1955} IMF by dividing by a factor of 1.8. We note that the \cite{jones2013} data are for gravitationally lensed galaxies and therefore the metallicity gradients may be subject to the uncertainties in reconstructing the galaxy images, although they have the advantage of being at high spatial resolution. We perform an outlier-resistant linear regression to the combined high redshift sample of KMOS-HiZELS, \cite{queyrel2012,jones2013} and \cite{swinbank2012}, which has the form $\frac{\Delta {\rm Z}}{\Delta \it{r}}=a\,\log({\rm M_{\star}})+b$ where $a=-0.022\pm0.009$ and $b=0.22\pm0.03$ (i.e. the slope is $2.4\sigma$ from being flat). No significant correlations are found between metallicity gradient and SFR or effective radius, although we note that in the local Universe \cite{sanchez2014} do find a correlation with radius. 

From a simple physical perspective we might also expect a trend between metallicity gradient and the kinemetry parameter, K$_{\rm Tot}$, as this is a measure of how disturbed the system is, or sSFR, as this is a measure of how intensely the galaxy is forming stars, both of which will be associated with the motion of gas within the galaxy. {Interestingly, we find no trend with K$_{\rm Tot}$ for our sample but this is perhaps not surprising as the galaxies are selected to be typical of $z\sim1$ star forming galaxies, for which the incidence of mergers is only $\sim10\%$ \citep{stott2013}, and probe only a small range in K$_{\rm Tot}$. If we combine our data with the K$_{\rm Tot}$ values measured in \cite{swinbank2012} there is still no trend, but again these are typical galaxies mainly in the K$_{\rm Tot}<0.5$ regime. Unfortunately, we do not have the K$_{\rm Tot}$ values for the rest of the literature data. }

Also displayed in Fig. \ref{fig:metgrad} is the metallicity gradient plotted against sSFR for which we do see a trend, which is strengthened when our data are combined with those of \cite{queyrel2012,jones2013} and \cite{swinbank2012}. As above, we perform a fit to this combined high redshift sample, which has the form $\frac{\Delta {\rm Z}}{\Delta \it{r}}=c\,\log({\rm sSFR})+d$ where $c=0.020\pm0.007$ and $d=0.18\pm0.07$ (i.e. the slope is $2.9\sigma$ from being flat). We note that the two \cite{jones2013} galaxies not shown in Fig. \ref{fig:metgrad} are significant outliers with metallicity gradients of $\sim-0.25\rm \,dex \,kpc^{-1}$ and $\rm sSFR\sim 3\times10^{-8}$ and $\rm \sim5\times10^{-9} yr^{-1}$.

For comparison with local galaxies we include data points inferred from \cite{rupke2010b} who study a sample of normal and merging star-forming galaxies in the local Universe. We derive SFRs for the \cite{rupke2010b} sample by using their tabulated far-infrared luminosities, assuming \cite{kennicutt1998} and note that the majority of the mergers have $\rm L_{IR}\gtrsim10^{10.5} L_{\odot}$, with three galaxies being luminous infrared galaxies (LIRGs, $\rm L_{IR}\gtrsim10^{11} L_{\odot}$). The majority of the non-mergers have ($\rm L_{IR}\lesssim10^{10} L_{\odot}$). We derive stellar masses for their galaxies using their tabulated absolute $K$ band magnitudes, assuming an underlying simple stellar population model from \cite{bc2003} that has a solar metallicity and was formed at $z=1$ (i.e. $\sim8$\,Gyr old). If we include the $z\sim0$ \cite{rupke2010b} sample in the sSFR, metallicity gradient fit then the significance of the trend increases with the parameters becoming $c=0.023\pm0.004$ and $d=0.20\pm0.05$ (i.e. the slope is $5.8\sigma$ from being flat). As stated above, the effect of any unseen, low level AGN activity would act to steepen the negative metallicity gradients but flattening the positive ones. This would therefore not affect the general trend of our result. We discuss the theoretical implications of the relationship between sSFR and metallicity gradient in \S\ref{sec:disc}. We also note that if we include the \cite{rupke2010b} sample in the metallicity gradient mass fit then the significance of that trend increases to only $2.7\sigma$ from a flat relation. 

\begin{figure}
   \centering
\includegraphics[scale=0.5, trim=0 0 0 0, clip=true]{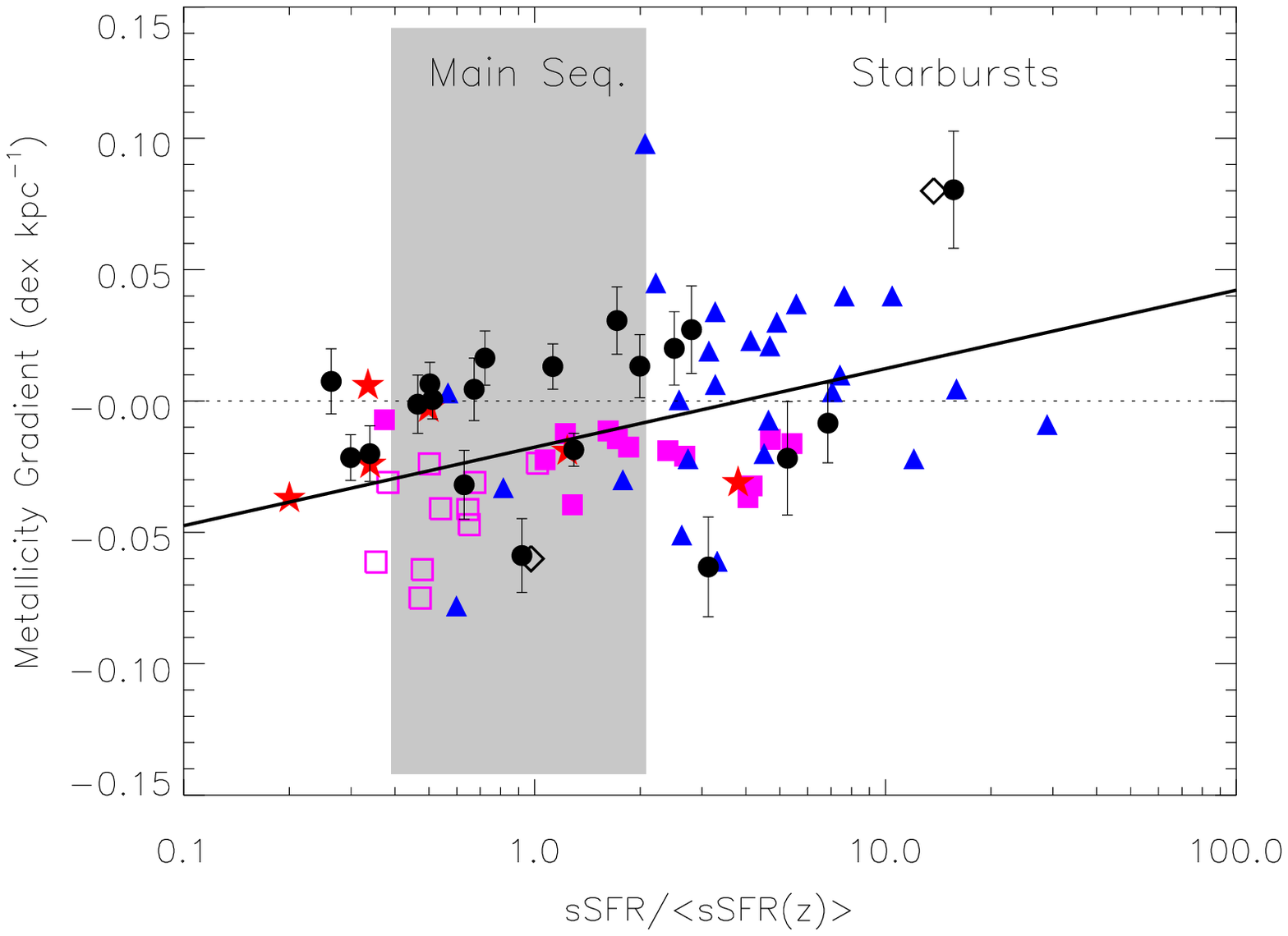}  
\caption[]{The metallicity gradient plotted against epoch normalised sSFR (sSFR$\rm _{EN}$=sSFR/$<$sSFR($z$)$>$) for the samples shown in Fig. \ref{fig:metgrad}. The grey region represents the main sequence (sSFR$\rm _{EN}=1$) with a factor of two in sSFR$\rm _{EN}$ either side. The starburst galaxies populate the region with sSFR$\rm _{EN}>2$. The solid line is a fit to the combined data of KMOS-HiZELS,  \cite{swinbank2012, queyrel2012} and \cite{rupke2010b}, which demonstrates that the starbursts tend to have more positive metallicity gradients.} 
   \label{fig:metgradms}
\end{figure}

Finally, the average sSFR of the main sequence of typical star forming galaxies increases with redshift \citep{elbaz2011,sobral2014}. Galaxies above this main sequence at a given redshift are often classed as `starbursts', i.e. those that are most vigorously forming stars at that epoch. This means that a galaxy classed as a starburst at low redshift will have the same sSFR as the typical main sequence galaxies at higher redshift. To account for this evolution, in order to compare main sequence galaxies to starbursts across all redshifts, we normalise the sSFR of the galaxies in Fig. \ref{fig:metgrad} by the average sSFR of the main sequence at their redshift (using \citealt{elbaz2011}). In Fig. \ref{fig:metgradms} we plot this epoch normalised sSFR (sSFR$\rm _{EN}$=sSFR/$<$sSFR($z$)$>$, c.f. epoch normalised SFR, \citealt{stott2013}) against metallicity gradient. A similar trend to that in Fig. \ref{fig:metgrad} is found, with the fit $\frac{\Delta {\rm Z}}{\Delta \it{r}}=e \,\log({\rm sSFR\rm _{EN}})+f$ returning a slope of $e=0.028\pm0.007$ ($f=-0.02\pm0.01$). The galaxies on the main sequence are found to have an average metallicity gradient of $-0.020\pm0.004$ while those in the starburst region have an average of $0.004\pm0.006$, a difference of $3.1\sigma$. We note from \cite{stott2013}, that galaxies on the main sequence at any redshift have a low major merger fraction ($\sim10\%$) whereas those in the starburst region may have a merger fraction of $\sim50\%$. {Environmental classifications for isolated and potentially interacting galaxies exist for the \cite{queyrel2012} sample (see \citealt{epinat2012}). From this we find that the interacting galaxies do have a higher biweight average metallicity gradient of $0.028\pm0.010\,\rm dex\,kpc^{-1}$ than the isolated galaxies, which have an average gradient of $0.003\pm0.008\,\rm dex\,kpc^{-1}$, but this is only a $\sim2\sigma$ difference. We note that this difference is reduced to $\sim1\sigma$ if we include the K$_{\rm Tot}>0.5$ defined mergers from KMOS-HiZELS and \cite{swinbank2012}. We discuss the implications of the sSFR$\rm _{EN}$ and merging for the metallicity gradients in \S\ref{sec:disc}.}

%SV2 ifu11 ifu14 ifu15 ifu16 ifu17 ifu18 ifu1 ifu23 ifu3 ifu4 ifu6 ifu8 ifu9

\begin{table}
\begin{center}
\caption[]{The metallicity gradients for the 20 KMOS-HiZELS galaxies where it was possible to measure them.}

\label{tab:massmet}
\small\begin{tabular}{lrr}
%\hline
\hline
\multicolumn{1}{c}{Galaxy}&\multicolumn{1}{c}{$\rm \frac{\Delta Z}{\Delta \it{r}}$} &$\rm \log(sSFR)$ \\
&\multicolumn{1}{c}{(dex\,kpc$^{-1}$)}&(yr$^{-1}$) \\
\hline
CFHT-NBJ-C339&$0.004\pm0.012$&-9.6\\
CFHT-NBJ-C343&$-0.020\pm0.011$&-9.9\\
CFHT-NBJ-956&$-0.059\pm0.015$&-9.4\\
CFHT-NBJ-1209&$-0.022\pm0.022$&-8.7\\

CFHT-NBJ-1709&$0.007\pm0.012$&-10.0\\
CFHT-NBJ-1739&$-0.001\pm0.011$&-9.7\\
CFHT-NBJ-1740&$0.016\pm0.010$&-9.5\\

CFHT-NBJ-1745&$0.025\pm0.017$&-9.0\\
CFHT-NBJ-1759&$-0.018\pm0.006$&-9.3\\
CFHT-NBJ-1774&$0.013\pm0.012$&-9.1\\
CFHT-NBJ-1787&$0.007\pm0.008$&-9.7\\

CFHT-NBJ-1789&$0.000\pm0.007$&-9.7\\
CFHT-NBJ-1790&$0.032\pm0.012$&-9.2\\
CFHT-NBJ-1793&$0.012\pm0.009$&-9.3\\
CFHT-NBJ-1795&$-0.063\pm0.019$&-8.9\\
CFHT-NBJ-2044&$-0.020\pm0.008$&-9.9\\
CFHT-NBJ-2048&$0.073\pm0.020$&-8.2\\

VVDS-503&$-0.010\pm0.015$&-8.6\\
VVDS-588&$-0.031\pm0.013$&-9.6\\
VVDS-888&$0.020\pm0.014$&-9.0\\

\hline
\end{tabular}
\end{center}
\end{table}

\subsubsection{The effect of atmospheric seeing and inclination}
\label{sec:see}

The metallicity gradient will be affected by the seeing. We measure the metallicity in elliptical annuli and expect galaxies with the largest ellipticities, due to their large angles of inclination, to be affected more than those that are face on. This is because the annuli are closer together in the minor axis direction and are separated by less than the HWHM of the seeing disc. 

We test the combined effect of seeing and inclination by performing a simulation of 1000 discs with random input metallicity gradients in the range $-0.2<\frac{\Delta {\rm Z}}{\Delta \it{r}}<0.2$ and random inclination angles of $0<i<90^{\circ}$. These discs are then smoothed with a Gaussian kernel, with a FWHM the same as the observed atmospheric seeing ($0.7''$). The results of this test are that for a face on disc, the metallicity gradient we observe in $0.7''$ seeing will be $\sim80\%$ of its true value. When we consider the inclination angle as well, the observed metallicity gradient at the median inclination of the KMOS-HiZELS sample ($50^{\circ}$) will only be $70\%$ of its intrinsic value. The most extreme correction to our sample comes for the galaxy with the largest positive gradient (CFHT-NBJ-2048, $\frac{\Delta {\rm Z}}{\Delta \it{r}}=0.07$, $i=80^{\circ}$) for which we may only be observing 30\% of its true value. For all other galaxies in KMOS-HiZELS we observe at least 50\% of their intrinsic metallicity gradient. 

We note that the flattening effect we see is less dramatic than that seen by \cite{yuan2013} (see also \citealt{mast2014}), who find that the measured value of the metallicity gradient is only $\sim10-20\%$ of its true value for a simulation of a degraded face on disc, in similar seeing conditions to those seen here. The reason for this is that our annuli are evenly spaced such that the outer radius is not contaminated by higher metallicity material close to the galaxy core. \cite{yuan2013} use the annular radii: $\rm \lesssim1.5$, $\rm \sim1.5-3.5$ and $\rm \sim3.5-9\,kpc$ (c.f. our annuli $\rm <3$, $\rm 3-6$ and $\rm 6-9\,kpc$). Also, the effect of the seeing on their galaxy is more significant as it is at $z=1.49$, where the FWHM of $0.7''$ corresponds to $\rm 6\, kpc$, whereas at $z=0.8$ this is $\rm 5\, kpc$.

The local samples of \cite{rupke2010b}, and the high-resolution, adaptive optics observed samples of \cite{swinbank2012} and \cite{jones2013} will be significantly less affected by seeing. However, the \cite{queyrel2012} sample will be affected in a similar way to our own. We therefore perform a seeing correction to our data using the individual inclination angle values and one to the \cite{queyrel2012} sample, assuming the average inclination of our sample for all of their galaxies. We refit the trend between sSFR and metallicity gradient using the seeing-corrected KMOS-HiZELS and \cite{queyrel2012} data and find that the parameters from \S\ref{sec:metgrad} become $c=0.026\pm0.006$ and $d=0.23\pm0.06$. These parameters represent an increased slope as expected but agree with the previous values within $1\sigma$. We therefore conclude that removing the effect of the atmospheric seeing only acts to strengthen our result.

\section{Discussion}
\label{sec:disc}

Simulations of the evolution of disc galaxies that predict inside-out growth tend to have initially steep negative abundance gradients which then flatten at later times (e.g. \citealt{marcon2010,stinson2010,gibson2013}). Alternatively, flattened and positive gradients have been interpreted as suggesting an inflow of metal-poor gas to their central regions. This may be triggered by either mergers \citep{rupke2010a} or cold flows \citep{keres2005,dekel2009,cresci2010}. The significant correlation we find between metallicity gradient and sSFR is consistent with this picture, as galaxies with an increased sSFR are thought to be fuelled by gas flowing towards their centres, caused by either merging or efficient accretion. 

\begin{figure}
   \centering
\includegraphics[scale=0.5, trim=0 0 20 0, clip=true]{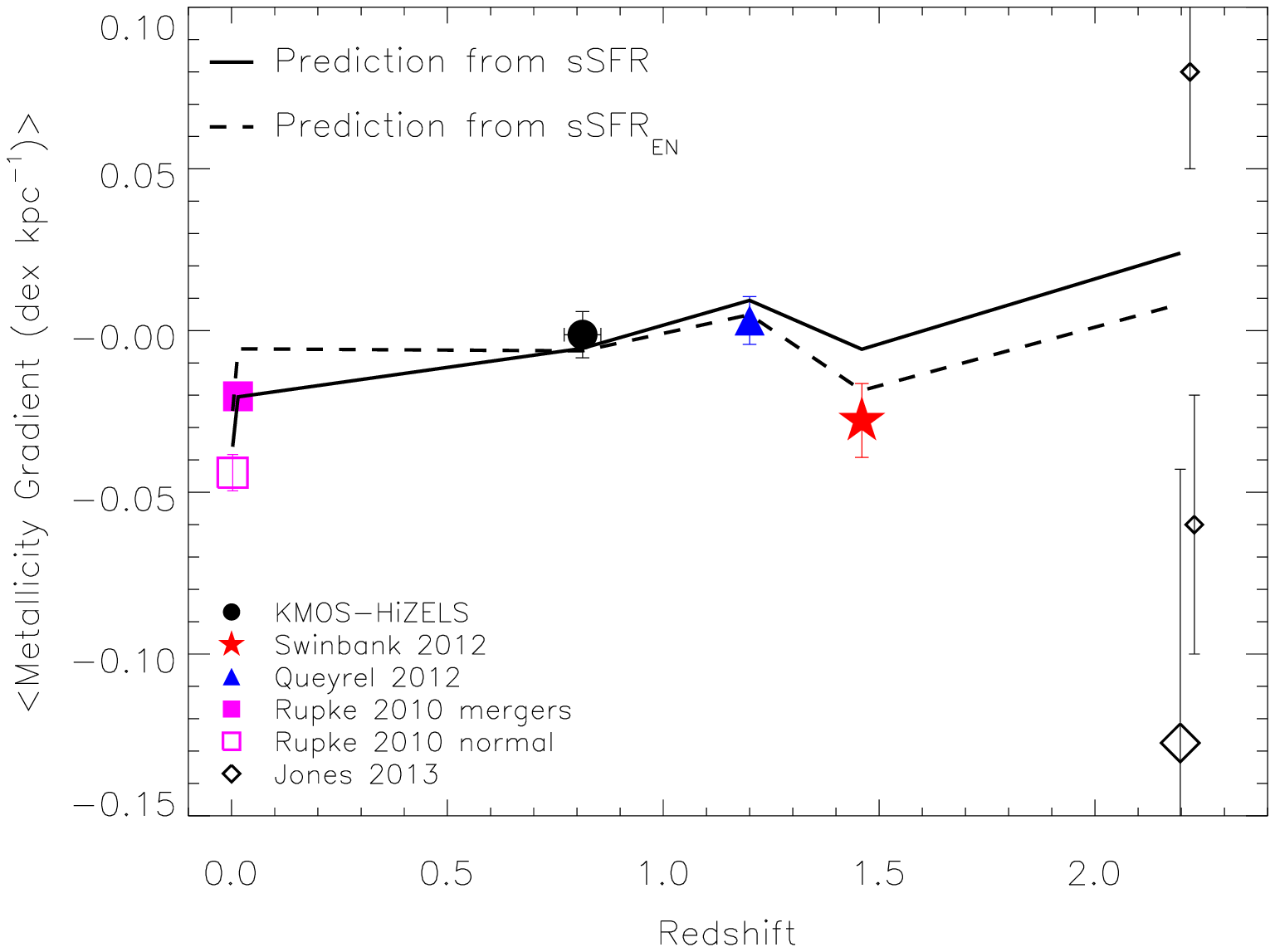}  
\caption[]{ The average metallicity gradient plotted against average redshift for the samples used in Fig. \ref{fig:metgrad}. Our KMOS-HiZELS and the \cite{queyrel2012} results indicates that the metallicity gradient becomes more positive with redshift whereas the \cite{swinbank2012} data suggests no evolution. Finally the \cite{jones2013} average, represented by the large open diamond, would suggest that the slope becomes more negative. We note that there are only four independent \cite{jones2013} galaxies and therefore the two galaxies with metallicity gradients of $-0.25\rm \,dex \,kpc^{-1}$ and $\rm sSFR\sim 3\times10^{-8}$ and $\rm \sim5\times^{-9} yr^{-1}$ (which were outliers to the fit in Fig. \ref{fig:metgrad}) lower the average significantly. However, the two smaller diamonds represent the other two galaxies in the \cite{jones2013} sample that do follow our trend. The solid and dashed lines are the prediction due to the trends between sSFR and sSFR$_{\rm EN}$ with metallicity gradient, respectively, as shown in Fig. \ref{fig:metgrad} and \ref{fig:metgradms}. They both successfully explain the competing claims for how the metallicity gradient appears to evolve with redshift. } 
   \label{fig:metgradz}
\end{figure}

Our results may explain why there are competing claims on how the gas phase metallicity gradient evolves with redshift (see Fig. \ref{fig:metgradz}) with some claiming positive gradients at high-$z$ \citep{cresci2010,queyrel2012} and others negative (e.g. \citealt{swinbank2012}) as it appears that this may just be driven by the different sSFR of the observed samples. For example, the median $\rm \log\,(sSFR\,yr^{-1})=-8.5$ for \cite{queyrel2012} who find a median metallicity gradient of $\rm +0.005\,dex\,kpc^{-1}$ and the median $\rm \log\,(sSFR\,yr^{-1})=-9.5$ for \cite{swinbank2012} who find an median metallicity gradient of $\rm -0.024\,dex\,kpc^{-1}$. It also explains the difference in slope between the `normal' star forming galaxies and merging LIRG-like systems, as seen by \cite{rupke2010b} in the local Universe. From a galaxy evolution perspective our findings mean that a galaxy's sSFR is governed by the amount of (typically metal poor) gas that can be funnelled into its core, triggered either by merging or efficient accretion. 

This picture is also in agreement with the observed fundamental metallicity relation (FMR, \citealt{mannucci2010,laralopez2010,stott2013b}) in which galaxies at a fixed stellar mass are found to be more metal poor with increasing SFR. In fact, as measurements of the FMR tend to use spectroscopy of the bright inner regions of the galaxy due to the limited size of spectroscopic slits and fibres (e.g. SDSS, \citealt{mannucci2010}; FMOS, \citealt{stott2013b}), then the relation between sSFR and metallicity gradient presented in this paper may help to explain the FMR. To quantify this we consider two galaxies, both of mass $\rm 1\times10^{10}M_{\odot}$, with $\rm SFR=1$ and $50\rm \,M_{\odot} yr^{-1}$ respectively. If we assume both galaxies have a solar metallicity (8.66 dex) at a galactocentric radius of $\rm 5\,kpc$ then we can use the relationship between sSFR and metallicity gradient to calculate their central metallicities. These are predicted to be 8.8\,dex and 8.6\,dex respectively, i.e. a difference of 0.2\,dex. Using the FMR equation of \cite{mannucci2010} the predicted difference in metallicity due to the difference in SFR at fixed mass is 0.25\,dex, which is in good agreement.

We now discuss whether we can determine if mergers or accretion are responsible for the trend between metallicity gradient and sSFR. The average sSFR of the main sequence of typical star forming galaxies increases with redshift \citep{elbaz2011}. In the redshift range $z\leq2.2$ the major merger fraction on the main sequence is found to be constant at $\sim10\%$ \citep{stott2013}, which suggests that the reason for the increase in sSFR is either secular processes, such as cold flows, or minor merging. However, the major merger fraction is found to increase with sSFR at a given redshift (i.e. relative to the star forming main sequence of that epoch), such that $\sim50\%$ of starbursts at any epoch are major mergers \citep{stott2013}. When these two observations are taken in concert it means that although a low redshift starburst galaxy is likely to be driven by a merger, it will have the same sSFR as the non-merging main sequence population at high redshift. Both the secular sSFR evolution of the main sequence and the increase in sSFR at a given epoch due to merging could provide a mechanism for forcing metal poor gas towards the centres of star forming galaxies. It is therefore difficult to separate out the effects of merging and secular processes as the cause for the trend in metallicity gradient with sSFR shown in Fig. \ref{fig:metgrad}. To account for this we normalise the sSFR to the average of the main sequence at the galaxies' redshift and plot this against metallicity gradient in Fig. \ref{fig:metgradms}. From this analysis we find that the average metallicity gradient of the main sequence galaxies is significantly more negative than the starbursts. This suggests that merging may play a significant role in driving the metallicity gradient to more positive values. In Fig. \ref{fig:metgradz} we include a prediction for the metallicity gradient of the samples with redshift based on the trend with the epoch normalised sSFR (sSFR$\rm _{EN}$) from \S\ref{sec:metgrad} and Fig. \ref{fig:metgradms}. {However, we note that the evidence for more positive metallicity gradients in galaxies with dynamical or visual indicators of merging is of low statistical significance ($\sim1-2\sigma$)}. Therefore, secular processes, which increase in efficiency with redshift, may still be important.

\section{Summary}
\label{sec:conc}
We have observed a sample of 39 typical star-forming galaxies at $z\sim1$ with KMOS to investigate their dynamics and metallicity gradients. From these data we conclude: 
\begin{itemize}
\item  The majority of the KMOS-HiZELS sample of $z\sim1$ star forming galaxies show disc-like rotation.
\item  The metallicity gradients of the galaxies are generally consistent with being either flat or negative (higher metallicity in the galaxy core relative to outer regions).
\item  There is a trend between sSFR and metallicity gradient, in that galaxies with a higher sSFR tend to have a relatively metal-poor centre. 
 \item  When we account for the average sSFR of the star forming main sequence it seems that the starbursts have significantly more positive metallicity gradients than typical galaxies.

\end{itemize}

The trend between sSFR and metallicity gradient suggests that the funnelling of metal-poor gas into the centres of galaxies, triggered via either merging or efficient accretion, is the driver of high sSFRs. In fact merging may play a significant role as it is the starburst galaxies at all epochs, which have the more positive metallicity gradients. The trend with sSFR helps to explain the conflicting observational claims for how the metallicity gradient of galaxies evolves with redshift.  Our results may also explain the FMR in which there is observed to be a negative correlation between metallicity and SFR at fixed galaxy mass.

\vspace{1in}
\noindent{\bf ACKNOWLEDGEMENTS}

Firstly, we acknowledge the referee for their comments, which have improved the clarity of this paper. JPS and IRS acknowledge support from STFC (ST/I001573/1). IRS also acknowledges support from the ERC Advanced Investigator programme DUSTYGAL and a Royal Society/Wolfson Merit Award. DS acknowledges financial support from NWO through a Veni fellowship and from FCT through the award of an FCT-IF starting grant. PNB acknowledges STFC for financial support.

\bibliographystyle{mn2e}
\bibliography{kmos}

\end{document}